\documentclass[aps,prb, twocolumn,amsmath,amssymb,longbibliography]{revtex4-1}
\usepackage{amsmath,amssymb,amsfonts,bm}
\usepackage{graphicx}
\usepackage{xcolor}
\usepackage{bbold}
\usepackage{graphicx}
\usepackage{dcolumn}
\usepackage{epstopdf}
\usepackage{epsfig}
\usepackage{url}
\usepackage{hyperref}
\usepackage{verbatim}
\usepackage[export]{adjustbox}
\usepackage{scalerel}
\usepackage{braket}
\newcommand{\Tr}{\operatorname{Tr}}

\newcommand{\be}{\begin{equation}}
\newcommand{\ee}{\end{equation}}
\newcommand{\ba}{\begin{aligned}}
\newcommand{\ea}{\end{aligned}}

\newcommand{\levc}[1]{K({#1})}

\def\OO{\mathcal{O}}
\def\GG{G}

\newcommand\Ddots{\scaleobj{0.55}{\ddots}}
\newcommand\Vdots{\scaleobj{0.55}{\vdots}}

\newcommand\Cdots{\scaleobj{0.55}{\cdots}}

\graphicspath{{figures/}}

\begin{document}

\title{Solution of a minimal model for many-body quantum chaos}
\author{Amos Chan,  Andrea De Luca and J. T. Chalker}
\affiliation{Theoretical Physics, Oxford University, 1 Keble Road, Oxford OX1 3NP, United Kingdom}

\date{\today}

\begin{abstract}
We solve a minimal model for an ergodic phase in a spatially extended quantum many-body system. The model consists of a chain of sites with nearest-neighbour coupling under Floquet time evolution. Quantum states at each site span a $q$-dimensional Hilbert space and time evolution for a pair of sites is generated by a $q^2\times q^2$ random unitary matrix. The Floquet operator is specified by a quantum circuit of depth two, in which each site is coupled to its neighbour on one side during the first half of the evolution period, and to its neighbour on the other side during the second half of the period. We show how dynamical behaviour averaged over realisations of the random matrices can be evaluated using diagrammatic techniques, and how this approach leads to exact expressions in the large-$q$ limit. We give results for the spectral form factor, relaxation of local observables, bipartite entanglement growth and operator spreading.
\end{abstract}

\maketitle
\tableofcontents

\section{Introduction} \label{intro}

Random matrix theory plays a central role in the understanding of chaotic quantum systems. \cite{Mehta} It is founded on the idea that for many systems of interest there is no privileged basis in Hilbert space. Some important phenomena in this field, however, arise only if there is spatial structure and a notion of locality. 
Diffusive transport in weakly disordered conductors is such an example for single-particle systems, and spreading of quantum information is a counterpart for many-body systems. It is natural to attempt to combine the simplifying features of random matrix theory with extended spatial structure. For diffusive conductors this is achieved in Wegner's $n$-orbital model, \cite{Wegner1979} in which hopping between sites of a tight-binding system is governed by $n\times n$ random matrices and disorder-averaged properties can be calculated exactly in the limit $n\to\infty$. Our aim in this paper is to establish a comparable simplification for spatially extended many-body systems.

Chaotic many-body quantum systems lie at the focus of efforts to understand the foundations of quantum statistical mechanics. \cite{Deutsch,Srednicki,Rigol,Polkovnikov} Generic features expected in the dynamics of such systems include rapid equilibration of local observables \cite{Rigol} and ballistic propagation of quantum information, \cite{Lieb} as well as ballistic growth of bipartite entanglement. \cite{KimHuse,HoAbanin} Conservation laws play a central part in dynamics, and one expects that systems with a given set of conservation laws will form a distinct class. Our focus in the following is on evolution arising from a time-dependent Hamiltonian: without even energy as a conserved quantity, it constitutes a particularly simple example.

Random matrix approaches offer natural routes to constructing models with minimal structure, and unitary quantum circuits provide an attractive way to formulate the evolution operator for time-dependent quantum systems. Unitary circuits that are random in both space and time have recently yielded valuable insights into chaotic quantum dynamics. \cite{Nahum2017,Nahum2017a,vonKeyserlingk2017, Hamma2012, *hamma2012ensembles, *zanardi2014local} Here we initiate an analytic study of unitary circuits that are random in space but periodic in time. 

We study a Floquet operator acting on a one-dimensional system consisting of $q$-state `spins' at each site. The Floquet operator is constructed from unitary matrices that couple adjacent sites. These $q^2\times q^2$ matrices are drawn independently from the circular unitary ensemble (CUE) and we compute physical properties averaged over the ensemble. Our key simplification is to treat the large-$q$ limit. We show that quantum dynamics in this system exhibits a range of features that are expected to be characteristic of ergodic many-body quantum systems: correlators of local observables decay rapidly in time and quantum information spreads ballistically, in the sense that the bipartite entanglement of an initial product state grows linearly in time and the out-of-time-order correlator \cite{LarkinOvchinnikov,MSS} (OTOC) shows the `butterfly' effect. Our approach also provides access to the spectral properties of the Floquet operator and we show (on scales much larger than the level spacing) that its eigenvalue correlations are those of the CUE. At a technical level, our calculations are based on the application of diagrammatic techniques developed previously for single-particle problems in mesoscopic physics involving random matrices from the CUE. \cite{Brouwer1996}

Some features of the large-$q$ limit are non-generic: as for random unitary circuits \cite{Nahum2017a,vonKeyserlingk2017} at large $q$, the distinction between the velocities associated with entanglement and operator spreading vanishes; and sublinear growth of entanglement, expected at long times in one dimensional disordered systems because of weak links, \cite{Nahum2017b} is absent.

 The balance of this paper is organised as follows. In Sec.~\ref{results}, we define the model, observables and results. In Sec.~\ref{diagrep}, we develop the diagrammatic approach for taking the ensemble average of a given observable. In Sec.~\ref{diagres}, we sketch the proof of the results presented in Sec.~\ref{results} using the tools developed in Sec.~\ref{diagrep}. Lastly, in Sec.~\ref{sum}, we summarise. Technical details are described in a series of appendices.

\section{Model, observables and results}\label{results}

\begin{figure}[htb]
	\includegraphics[width=0.45\textwidth]{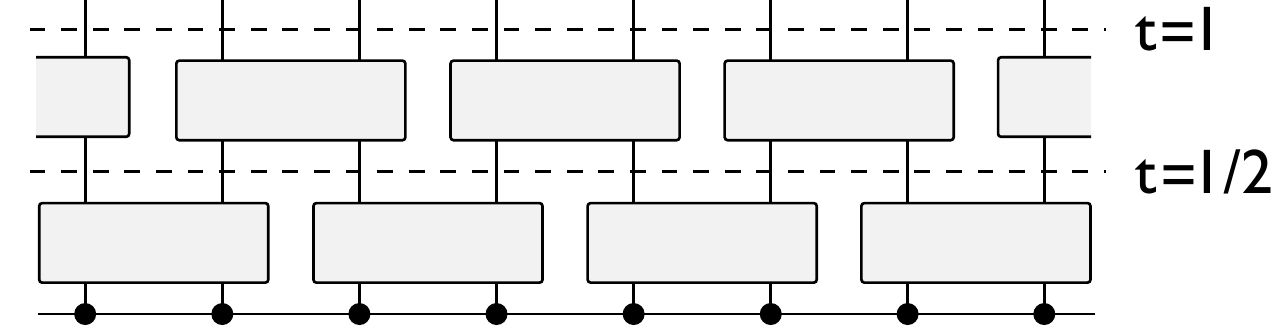}
	\caption{Illustration of the model for Floquet time evolution studied in this paper. Space and time are represented by the horizontal and vertical directions. Lattice sites are indicated by filled dots, and the coupling of pairs of sites under time evolution is shown with rectangles.}
	\label{fig:model}
\end{figure}

We seek a minimal model for quantum chaos in a spatially extended many-body system with local interactions. 
We formulate the model directly in terms of the time-evolution operator, rather than a Hamiltonian, and for simplicity we consider a Floquet problem. Taking a one-dimensional lattice, evolution over one period is separated into two half-steps. Each even site is coupled to its neighbour on the left in the first-half-step, and to its neighbour on the right in the second half-step. 
This quantum circuit is illustrated in Fig.~\ref{fig:model}.

To write the evolution operator explicitly, let $U_{2n-1,2n}$ denote the unitary matrix that couples sites $2n-1$ and $2n$ in the first half-step, and let $U_{2n,2n+1}$ be the matrix that couples sites $2n$ and $2n+1$ in the second half-step. Each $U_{i,i+1}$ is independently distributed with the Haar measure and we calculate physical properties as an average over this ensemble, denoted by $\langle \ldots \rangle$. For a system of $L$ sites the full Hilbert space has dimension $q^L$. Taking $L$ even for definiteness, the first half-step is represented by an evolution operator acting in this space with the form
\begin{equation}
W_1 = U_{1,2} \otimes U_{3,4} \otimes \ldots \otimes U_{L-1,L}\,.
\end{equation}
The evolution operator for the second half-step is similarly
\begin{equation}
W_2 = \openone_q \otimes U_{2,3} \otimes U_{4,5} \ldots U_{L-2,L-1} \otimes \openone_q\,,
\end{equation}
where $\openone_q$ denotes the $q\times q$ unit matrix. The Floquet operator describing evolution over a single complete period is
\begin{equation}
W = W_2 \cdot W_1\,.
\end{equation}
We denote the evolution operator for an integer number $t$ of periods (the $t$-th power of $W$) by $W(t)$.

To demonstrate that this model has the features expected in a chaotic quantum many-body system, we examine a range of physical properties, as detailed below. Our results for entanglement spreading and for the OTOC are the same as those for random unitary circuits with the same structure as our model, if one takes the large $q$ limit in previous results. \cite{Nahum2017a,vonKeyserlingk2017} In this context, one of our significant conclusions is that random unitary circuits do indeed share important physical features with deterministic systems. At the same time, our model opens up discussion of the spectral properties of the evolution operator. This has no equivalent for random unitary circuits since they lack any such fixed operator.


\subsection{Spectral form factor}

 Our results for the spectral form factor show that the Floquet operator for the model has exactly the same eigenvalue correlations in the large $q$ limit as an ensemble of Haar-distributed unitary matrices. This insensitivity to the spatial structure of the system is a striking emergent feature.

The spectral form factor $K(t)$ is the Fourier transform of the two-point correlation function of the eigenvalue density. Denote the eigenphases of $W$ by $\{\theta_m\}$ for $m=1,\, \ldots q^L$. Then  
\be
\ba
K(t) \equiv \langle  \Tr [W(t)] \Tr [W^\dagger(t)] \rangle
= \sum_{m,n} \langle e^{i (\theta_m - \theta_n) t} \rangle \,.
\ea
\ee
For $N\times N$ matrices from the CUE one has \cite{Mehta}
\be
\label{Ktdef}
\levc{t} = \left\{\begin{array}{ccl}
	N^2 & \quad &t=0\\
	|t| & & 0< |t| \leq N\\
	N & & N \leq |t| \,.\end{array}\right.
\ee
The behaviour of $K(t)$ on scales $|t| \ll N$ reflects level correlations at separations much larger than the mean spacing, and the linear dependence of $K(t)$ on $|t|$ in this regime is a consequence of Coulombic suppression of long wavelength fluctuations in the eigenvalue density. Conversely, the form of $K(t)$ for $|t|\sim N$ encodes spectral correlations on the scale of the level spacing.

We obtain exactly the CUE form, $K(t)=|t|$ for $t\not=0$, in the large $q$ limit at fixed $L$ and $t$. We stress that this result is a consequence of coupling between all sites, and is not simply a reflection of the properties of individual random matrices $U_{i,i+1}$. To illustrate the point, consider an alternative model in which the couplings $U_{2n,2n+1}$ on even bonds are replaced with unit matrices $\openone_{q^2}$. For this toy system of $L/2$ independent pairs of sites one has much larger fluctuations in eigenvalue density, with $K(t) = |t|^{L/2}$ for $0<|t| \leq q^2$. 

Returning to the original model, it is interesting to speculate on behaviour in regimes other than the one we are able to analyse exactly. At large but fixed $q$ and fixed $t$, it is natural to expect $K(t)$ to increase with $L$ if $L$ is sufficiently large, since distant parts of the system should be only weakly correlated; results for this regime will be presented elsewhere. \cite{future}  In the opposite limit, fixing $q$ and $L$ and varying $t$, the form of $K(t)$ at $|t| \sim q^L$ probes correlations on the scale of the level spacing. We expect these to be of CUE form for all $q$ and $L$, provided only that $q^L \gg 1$. They should therefore show a transition from $K(t) = |t|$ for $|t|\lesssim q^L$ to $K(t) = q^L$ for $|t| \geq q^L$. The transition to constant $K(t)$ at large $t$ is, however, well-known to be inaccessible in a perturbation expansion in inverse powers of $q$.\cite{Efetov,Zirnbauer1996}

\subsection{Dynamics}

We next examine physical properties chosen to reveal both the dynamics of local degrees of freedom and the spreading of quantum information.

\subsubsection{Local relaxation}

A characteristic feature of chaotic dynamics is that local observables relax rapidly towards equilibrium. To probe relaxation in our model we introduce an operator $O_x$ representing an observable at site $x$ and obeying $O_x^\dagger = O_x$, ${\rm Tr}\, O_x = 0$ and $O_x^2 = \openone_q$ . Let
\be
O(x) = \openone_q \otimes \ldots \otimes \openone_q \otimes O_x \otimes \ldots \otimes \openone_q
\ee
and $O(x,t) = W^\dagger(t) O(x) W(t)$. The statistical average for a system with density matrix $\rho$ is $[ \ldots ]_{\rm av} = {\rm Tr}\,[ \rho \ldots]$. In our Floquet model the infinite-temperature density matrix $\rho = q^{-L} \openone_{q^L}$ is the natural choice and we denote the normalised many-body trace by ${\rm tr} \equiv q^{-L} {\rm Tr}$.

One anticipates that $[O(x,t) O(x)]_{\rm av}$ will relax to the value $[O(x)]_{\rm av}^2$ (which is zero with this form for $O(x)$) on a microscopic timescale of order the Floquet period. At large $q$ we in fact find complete relaxation within a single period,  obtaining for $q \to \infty$ the result
\be
\label{2pt}
\langle {\rm tr}[O(x,t) O(x)] \rangle= \delta_{t,0}\,.
\ee
We discuss the leading corrections to this result for finite $q$ and small $t$ in Sec.~\ref{local}.

\subsubsection{Entanglement spreading}

We probe entanglement spreading via the time dependence of the ensemble-averaged bipartite entanglement purity (and, more generally, ensemble averages of exponentials of R\'enyi entropies) for an initial state $|\psi\rangle$ that is a direct product over sites. Specifically, let $A$ be the left half of the system (with site labels $1\leq n \leq L/2$) and $B$ the right half, and consider the reduced density matrix
\be
\label{reddef}
\rho_A(t) = {\rm Tr}_B [ W(t) |\psi\rangle \langle \psi | W^\dagger(t)]\,.
\ee
The R\'enyi entropies $S_\alpha(t)$ are given by
\begin{equation}
\label{renyidef}
e^{ (1-\alpha) S_\alpha(t)}   =  \Tr_A \,[ \rho_A(t)]^\alpha  
\end{equation}
and the entanglement purity is ${\cal P}(t) = e^{-S_2(t)}$. We compute $\langle e^{ (1-\alpha) S_\alpha(t)}\rangle$ for $\alpha = 2$ and $\alpha =3$. 
The purity shows an exponential decay in time until it falls to a value ${\cal P}(t) \sim q^{-L/2}$ typical of random states in the many-body Hilbert space. In detail, we obtain at large $q$ 
\begin{equation} \label{reneq}
\langle e^{(1-\alpha) S_\alpha(t)}\rangle  \sim \left\{
\begin{array}{cll}
f_\alpha(t)\;	 q^{-2 (\alpha-1) t} & \quad & t\leq L/4 \\
K_\alpha q^{- (\alpha-1) L/2} & \quad & t > L /4
\end{array}\right.
\end{equation}
The function $f_\alpha(t)$ grows exponentially with $t$, and 
in particular, we find $f_2(t) = 4^t$ and $f_3(t) \simeq  ((4 + 3 \sqrt{2})/2)^{2t}$ for $t\gg 1$.
At times $t > L/4$, the Renyi entropies saturate and we conjecture that 
$K_\alpha = \text{Cat}(\alpha)$, where Cat is the $\alpha$-th catalan number. The latter result is proven for $\alpha$ up to 5 (see Appendix \ref{3renapp}). This is consistent with the generalized Page formula for bipartite entanglement in random states\cite{Page1993}\cite{Nadal2010} and with the results from random unitary circuit\cite{Zhou2018}.

An interpretation of these results is that the reduced density matrix for the pure initial state spreads at time $t$ over the Hilbert space spanned by basis states at $2t$ sites. As an illustration, suppose that $\rho_A(t)$ has $q^{2t}$ non-zero eigenvalues that are all equal. Then ${\rm Tr}_A \rho_A(t) = q^{-2(\alpha - 1)t}$, which is consistent with the leading order behaviour of (\ref{reneq}) for $t\leq L/4$. This demonstrates that the entanglement spreads ballistically with a velocity at large $q$ that is the same as the naive light-cone velocity, $v=2$, introduced below in Fig.~\ref{fig:model2}. The values of $f_2(t)$ and $f_3(t)$ give information on the distribution of non-zero eigenvalues of $\rho_A(t)$, and it is noteworthy that these two quantities are distinct.

The growth of entanglement found here is similar to that obtained for integrable systems using conformal field theories~\cite{calabrese2005evolution, *calabrese2007quantum, *calabrese2016quantum}. In integrable systems this behaviour is associated with the presence of quasiparticles that travel ballistically; a quite different physical picture is required for ergodic systems.

\subsubsection{Out-of-time-order correlator \label{OTOCsec}}

The spreading with time of local operators is characterised by the behaviour of the (ensemble-averaged) commutator
\be
\label{OTOCdef}
{\cal C}(x-y,t) = \frac{1}{2} \langle {\rm tr} |[O(x,t),O(y)]|^2 \rangle\,,
\ee
which measures how a local perturbation at site $x$ affects measurements at a later time at the site $y$. 
With our normalisation for local operators, one has
\be
\label{OTOCsimple}
{\cal C}(x-y,t)
= 1 - \langle {\rm tr} \left[ O(x,t)O(y)O(x,t)O(y) \right]\rangle
\ee
and the second term on the right-hand side is the OTOC. For short times $t$ and large separations $x-y$, the operators $O(x,t)$ and $O(y)$ commute and ${\cal C}(x-y,t)$ is zero, while for large times the OTOC is small and ${\cal C}(x-y,t)$ approaches unity. We obtain at large $q$ a sharp light-cone behaviour
\be\label{OTOC}
{\cal C}(x-y,t)= \left\{\begin{array}{lll}
	0&\quad & |t| < |x-y|/2\\
	&&\\
	1& & |t| \geq |x-y|/2\,.
\end{array}\right.
\ee
Hence operator spreading occurs with a butterfly velocity that, like the entanglement velocity, is equal to the naive light-cone velocity $v=2$ in the large $q$ limit. Results on random unitary circuits  \cite{Nahum2017a,vonKeyserlingk2017} suggest that all three velocities should be distinct for $q$ finite. From the same comparison, we also expect at finite $q$ that the step function of (\ref{OTOC}) will broaden.

\section{Ensemble averaging}\label{diagrep}

We now set out in several steps the general approach that we use to obtain these results. 

\begin{figure*}[t]
	\centering
	\includegraphics[angle=0, width=0.9\linewidth]{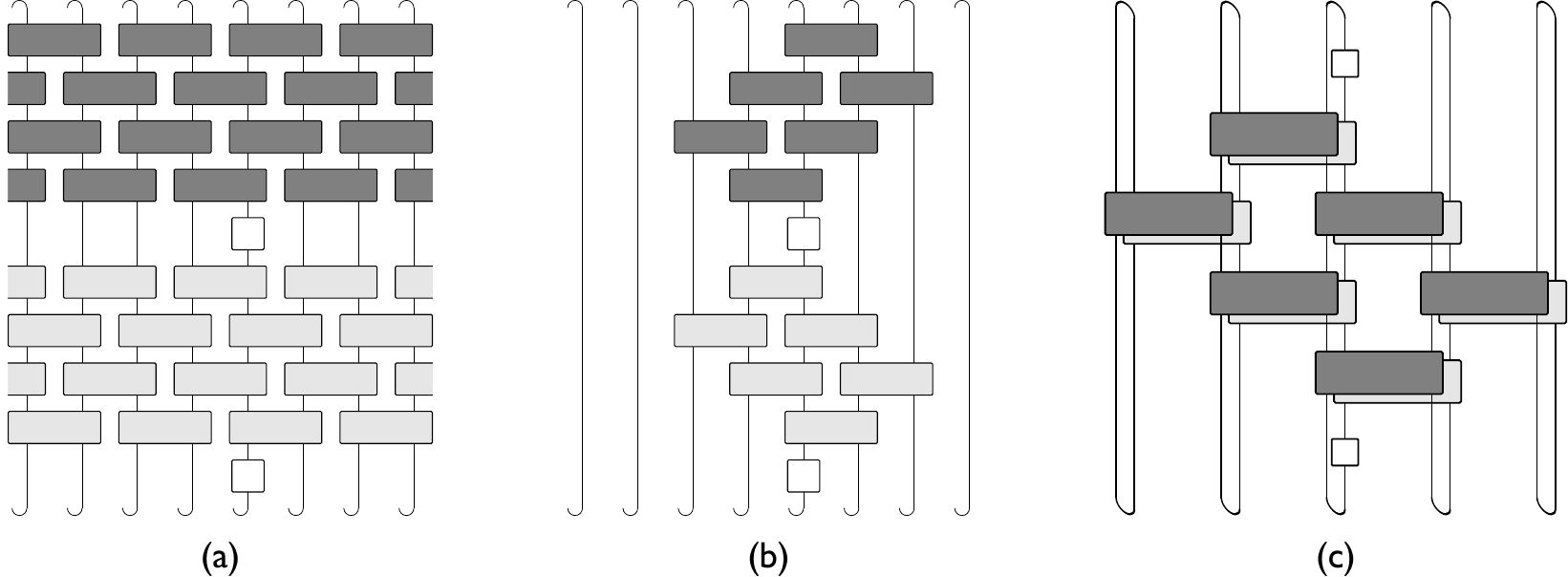}
	\caption{Illustration of $\Tr[O(t) O(0)]$ for $t=2$. $U$s and $U^\dagger$s are represented by rectangles, respectively shaded in light and dark grey. The local operator $O_x$ is represented by a square. The curled lines at the top and bottom edge indicate closed loops which denote the trace operation. (a): Initial expression. (b): After cancellation of pairs of $U$ with $U^\dagger$ wherever possible. The boundaries of the regions within which $U$s and $U^\dagger$s remain form lightcones with a velocity $v=2$ set by the construction of the model. 
(c): Illustration of $\Tr[O(t) O(0)]$ for $t=2$ formed by folding Fig.~\ref{fig:model2} (right) so that timelines run upwards for $W$ and downwards for $W^\dagger$. 
}\label{fig:model2}
\end{figure*}

First, it is useful to extend the notation of Fig.~\ref{fig:model} in various ways, so as to represent pictorially the quantities defined in Sec.~\ref{results}. An example for $\Tr[O(x,t) O(x,0)]$ is shown in Fig.~\ref{fig:model2}. The vertical timelines of sites join rectangles representing factors of $U_{i,i+1}$ and indicate matrix multiplication. Repeated copies of $W$ denote multiple time steps; $W^\dagger$ is shown as a differently shaded version of $W$; and local operators appear as squares. The matrix trace is shown by joining the timelines of sites to form closed loops. Finally, product states in the site basis (which do not appear in this example) are shown using circles at the ends of the timelines.  

A straightforward simplification in many instances is that some factors of $U$ and $U^\dagger$ cancel, as illustrated in Fig.~\ref{fig:model2}b, and the naive light-cone velocity $v=2$ clearly emerges. 

As a second step, it is helpful to fold the pictorial representations, so that while the timelines in $W$ run upwards, those in $W^\dagger$ run downwards, as shown in Fig.~\ref{fig:model2}c.

\begin{figure}[h]
	\centering
	\includegraphics[angle=0, width=0.9\linewidth]{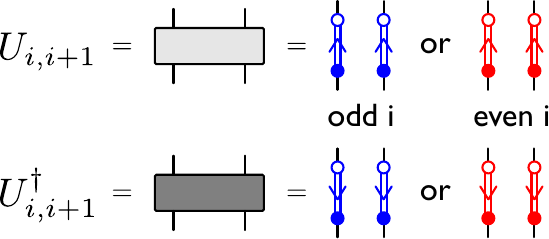}
	\caption{Adaptations to diagrammatic notation made in order to show time evolution of an individual site. The shaded rectangles representing $U_{i,i+1}$ and $U^\dagger_{i,i+1}$ are replaced by pairs of double lines, with an arrow directed from the column-label to the row-label of the matrix. To record the distinction between odd and even $i$, the double lines are blue in the first case and red in the second. This colour convention allows us to represent single-site diagrams without ambiguities (see e.g. Fig.~\ref{level1}).} 
	\label{diag1}
\end{figure}

These folded pictures provide a direct depiction of physical quantities but are cumbersome. A simpler representation is possible if we focus on the time evolution of a single site. To this end, and in anticipation of the disorder average, we switch to an alternative notation for $U_{i,i+1}$ and $U^\dagger_{i,i+1}$ in which individual sites appear separately, as shown in Fig.~\ref{diag1}.


With these preliminaries in place, we can set out a diagrammatic representation for ensemble averages.
It is a many-body extension of the one introduced for averages over the CUE by Brouwer and Beenakker. \cite{Brouwer1996} 
It can be applied to an arbitrary observable $f(W)$ but to be definite we illustrate it for $K(t)$ with $t=2$ and $L=2$, so that $W(2) = [U_{1,2}]^2$. In general three steps are involved.
	
\begin{figure}[h]
	\centering
	\includegraphics[angle=0, width=0.8\linewidth]{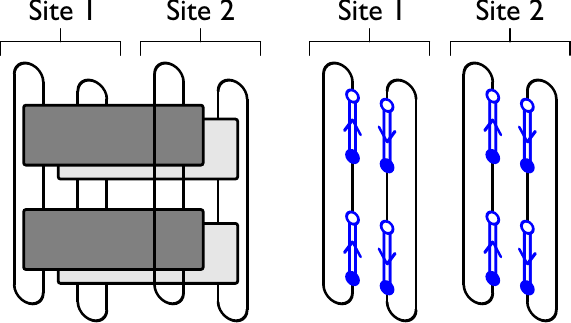}
	\caption{Alternative diagrammatic representations of ${\rm Tr}[W(t{=}2)] {\rm Tr}[W^\dagger(t{=}2)]$, which gives $K(t{=}2)$ after ensemble-averaging, using the notation of the left and right sides of Fig.~\ref{diag1}. 
	\label{ex1}
	}
\end{figure}

\begin{figure}[h]
	\centering
		\includegraphics[angle=0, width=1\linewidth]{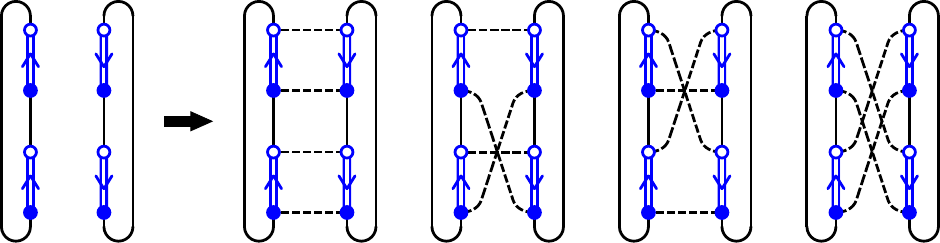}
        \caption{Single-site diagrams associated with ${\rm Tr}[W(t{=}2)] {\rm Tr}[W^\dagger(t{=}2)]$ (left) and  its ensemble average $\levc{t=2}$  (right). 
        }\label{ex2}
\end{figure}

	(i) The observable $f(W)$ is represented as a collection of single-site diagrams using the notation introduced in Fig.~\ref{diag1}.
{For the example of $\levc{t{=}2}$, one starts from the many-body diagram shown in Fig.~\ref{ex1} (right) and obtains the single-site representation shown in Fig.~\ref{ex2} (left). 
}
       %
       %
       
       (ii) The ensemble average $\langle f(W) \rangle$ is computed by generating a collection of single-site `contracted' diagrams $\GG = \{ G_i\}_{i=1}^L$ as follows. On each site, filled $U$-dots are connected to filled $U^\dagger$-dots of the same color on the same site with a dashed line (a contraction) in all possible ways, and likewise for the empty dots.  
       
       Since $U$s and $U^\dagger$s act on neighbouring pairs of sites, these contractions must be matched: for even $i$ the choice of contractions between blue dots must be the same in the diagrams $G_{i-1}$ and $G_i$; similarly for the red dots and the diagrams $G_i$, $G_{i+1}$. We refer to this as the \emph{bond constraint}.

\begin{figure}[h]
	\centering
	\includegraphics[angle=0, width=0.75\linewidth]{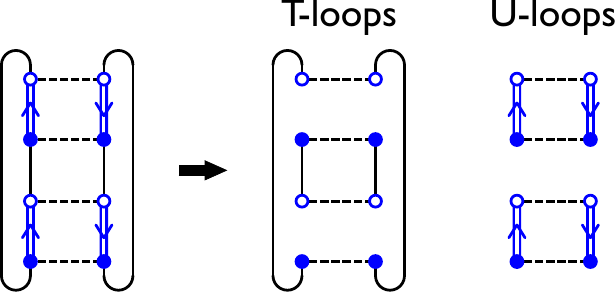}
	\caption{The $T$- and $U$- loops (right) associated with a contracted diagram for ${\rm Tr}[W(t{=}2)] {\rm Tr}[W^\dagger(t{=}2)]$  (left in this figure and top middle in Fig.~\ref{ex2}). The associated algebraic factors are $\mathcal{A}_T(G_i) = q^{2}$ and $\mathcal{A}_U(G_i) =(V_{1,1})^{1/2}$.}\label{tloopuloop}
\end{figure}

(iii) Each contracted site diagram $G_i$ gives rise to an algebraic expression $\mathcal{A}(G_i)$, obtained as the product of two factors
\begin{equation}
\mathcal{A}(G_i) = \mathcal{A}_T(G_i) \mathcal{A}_U(G_i) \;.
\end{equation}
These factors are associated with loops of two kinds, called $T$-loops and $U$-loops in Ref.~\onlinecite{Brouwer1996}.  The $T$-loops simply record pairings of matrix labels, and give rise to powers of $q$ in the contribution of a diagram. The $U$-loops distinguish different contributions from the average of factors of $U$ and $U^\dagger$. Examples of these two types of loop are illustrated in Fig.~\ref{tloopuloop}.

A $T$-loop is a closed sequence of alternating single and dashed lines. It carries an index corresponding to one of the $q$ basis states at a site. 
We generate the contribution associated with each $T$-loop by summing over its index. This leads to a factor of $q$ for a $T$-loop of single lines if it does not pass through any operator insertion, or of $\Tr (O_1 O_2 \dots )$ if the $T$-loop passes through the operators $\{ O_1, O_2, \ldots \}$. The overall factor $\mathcal{A}_T(G_i)$ is obtained as the product of the individual factors coming from each $T$-loop.

A $U$-loop is a closed sequence of alternating double and dashed lines. The length $c$ 
of a $U$-loop is defined as half of the number of double lines it contains. Let $R_i = \{c_k^{(r)}\}_{k=1}^{r_i}$ and $B_i = \{c_k^{(b)}\}_{k=1}^{b_i}$ be the sets of lengths of red and blue $U$-loops in $G_i$. Then from the theory of CUE averages \cite{Brouwer1996} 
	\begin{equation}
	\label{VVdef}
	\mathcal{A}_U(G_i)= (V_{R_i} V_{B_i})^{1/2}
	\end{equation}
	where the explicit form of the coefficients $V_{R_i} \equiv V_{c_1 \ldots c_{r_i}}$ and $V_{B_i} \equiv V_{c_1 \ldots c_{b_i}}$ (also known as the Weingarten functions) is given in App.~ \ref{recur}. Note that the exponent $1/2$ in (\ref{VVdef}) arises because we distribute the contribution arising from unitary operators equally over the two sites on which the operators act.

%
The final value of the ensemble average reads
\begin{equation}
\label{fWave}
\langle f(W) \rangle = \sum_{\GG} \prod_i \mathcal{A}(G_i)
\end{equation}
where the sum runs over all contracted diagrams $\GG$ satisfying the bond constraint. 
For the example of $\langle \levc{t{=}2}\rangle$, all possible single-site contracted diagrams are drawn in Fig.~\ref{ex2} right. 
Explicit evaluation of the algebraic expressions for this case leads to
\be 
\langle \levc{t=2} \rangle =  \left(q^4 V_{1,1} + q^2 V_2 + q^2 V_2 + q^4 V_{1,1} \right) \approx 2  \, .
\ee
The factors of $q$ within the bracket are contributions from the $T$-loops. We have taken the large-$q$ limit on the far right,  using $V_{1,1}= (q^4-1)^{-1}$ and $V_2 = -[q^2(q^4-1)]^{-1}$. 
At large $q$ the first and fourth diagrams are leading order, while the second and third are sub-leading. 

The above procedure is exact for any $q$, but the sum in \eqref{fWave} is problematic in general as it may involve an extremely large number of terms. For example, a total of $(t!)^{2(L-1)}$ diagrams contribute to $K(t)$. It is therefore useful
to establish which terms dominate in the large-$q$ limit. For large $q$ one has $\mathcal{A}_U(G_i) \sim q^{u_i - n_i}$ where $u_i = r_i + b_i$ is the total number of $U$-loops and $n_i=\sum_k c_k$ is the number of contractions on the $i$-th site. Let $\tau_i$ be the number of $T$-loops in $G_i$ that contribute a factor $q$. Then we have the large-$q$ expansion
\be \label{mborder}
\prod_i \mathcal{A}(G_i) \sim 
\prod_{i \in \text{site}} q^{\tau_i+u_i-n_i} \equiv \OO(\GG) 
\ee
where we have introduced the order $\mathcal{O}(G)$ and omitted a proportionality constant independent of $q$. In Appendix ~\ref{orderdet}, we discuss ways of enumerating the leading order diagrams. As a general rule, since the total number of contractions $\sum_i n_i$ is fixed, we should retain in \eqref{fWave} only the diagrams that maximize the total number of $T$- and $U$-loops. 

A natural approximation at large $q$ is to treat the elements of $U$ as independent Gaussian random variables, so that a standard Wick theorem applies. This approximation corresponds to including all diagrams where all $U$-loops are of unit length, and omitting all others. We refer to this set as the {Gaussian diagrams}. As an example, in Fig.~\ref{ex2}, the first and fourth diagrams on the right hand side are Gaussian. For these diagrams, Eq.~\eqref{mborder} holds with proportionality constant one; therefore counting the overall number of $U$- and $T$-loops is sufficient to obtain the leading contribution of the diagram.

In this paper, the leading contributions to all quantities calculated (except for the autocorrelation function of a local observable for $t>0$) are Gaussian. A procedure that goes beyond the Gaussian approximation is necessary, firstly, for the proof of this statement, and secondly, for future discussion of the sub-leading contributions. Since the number of diagrams contributing to each of the quantities we consider is finite at fixed $L$ and $t$, our results are exact in the limit $q\to \infty$ with $L$, $t$ fixed.

\section{Diagrammatic evaluation of results}\label{diagres}

We now show how this diagrammatic approach can be used to generate the results given in Sec~\ref{results}. We sketch the main ideas here, deferring formal proofs to appendices.

\subsection{Spectral form factor}\label{levelsec}

\begin{figure}[h]
	\centering
	\includegraphics[angle=0, width=1\linewidth]{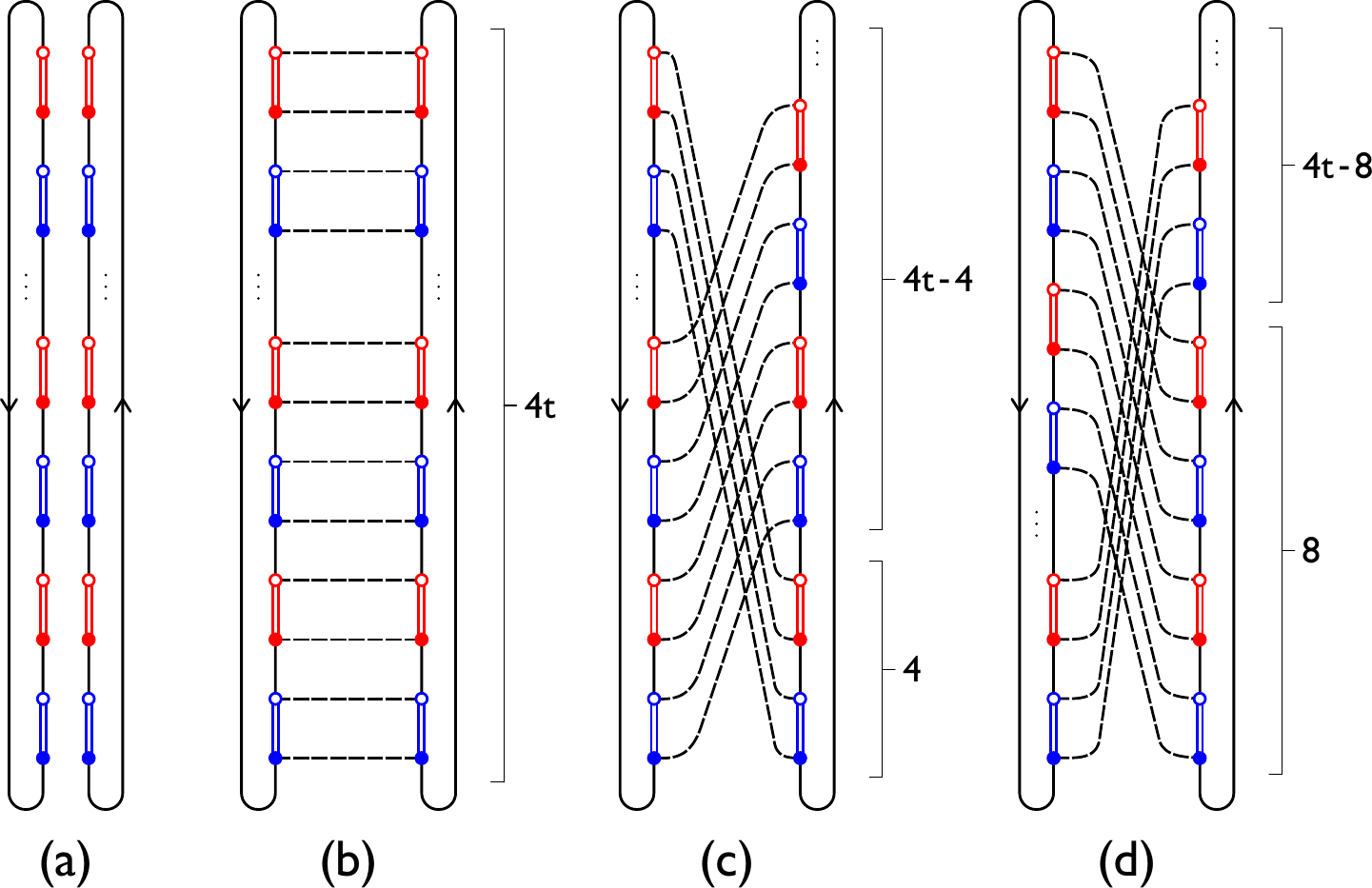}
	\caption{(a): The site diagram associated with  ${\rm Tr}[W(t)] {\rm Tr}[W^\dagger(t)]$. It is colour coded as follows: for an odd site $i$, the blue dots and double lines are contributions from $U^{}_{i,i+1}$ or $U^{\dagger}_{i,i+1}$ and the red ones are from $U^{}_{i-1,i}$ or $U^{\dagger}_{i-1,i}$. For an even site, this coding is reversed. 
		(b), (c) and (d):  Three examples from a total of $t$ leading order diagrams for $\levc{t}$ at $t >0$.
	}\label{level1}
\end{figure}

Consider the spectral form factor $\levc{t}$ introduced in \eqref{Ktdef}. Applying the procedure described in Sec.~\ref{diagrep} (and generalising the example given for $L=2$ and $t=2$ in Fig.~\ref{ex2}), we obtain for $L>2$  
a many-body diagram consisting of single-site diagrams as shown in Fig.~\ref{level1}a.
Here each timestep in $W(t)$ contributes with two unitaries, leading to $4t$ dots divided into four types -- blue or red, and empty or filled. 

Next we consider ensemble-averaging this diagram, making all possible contractions.
On a single site, one can easily check that one of the diagrams with the maximum number of loops is, for example, the site diagram in Fig.~\ref{level1}b.
Since this diagram is Gaussian and $\tau_i = u_i = 2t$ and $n_i = 4t$, its contribution for large $q$ is simply $1$. 
There are $t$ equivalent diagrams of this kind, obtained by cyclically shifting the right dots with respect to the left ones, as shown in Fig.~\ref{level1}c and Fig.~\ref{level1}d. 
Additionally, once one of these configurations has been chosen on the site $i$, the bond constraint forces all other sites to be in the same configuration in order to maximise the number of loops. All other diagrams are smaller by powers of $q$ for $q\to \infty$. In consequence we get the result $ \levc{t} = |t|$. A proof of this statement is given in Appendix \ref{levelcorr}.


\subsection{Relaxation of local observables}\label{local}

\begin{figure}[h]
	\centering
	\includegraphics[angle=0, width=0.8\linewidth]{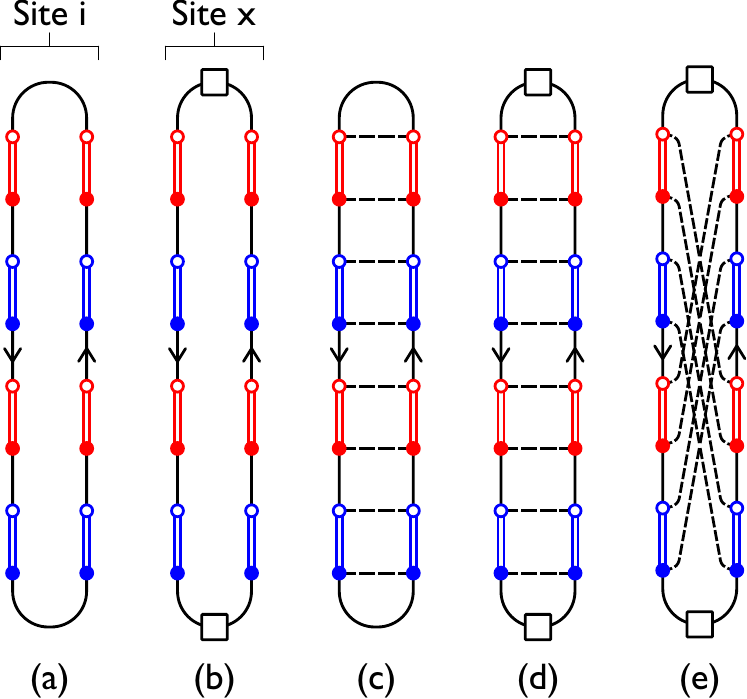}
	\caption{
		(a) and (b): Site diagrams associated with $[O(x,t{=}2) O(x)]_{\rm av}$ at sites $i \neq x$ and sites $i=x$ respectively. Color coding is as in Fig.~\ref{level1} .
		(c): The leading contraction of the site diagram for $x \neq i$.	(d) and (e): Two alternative contractions of the site diagram for $[O(x,t{=}2) O(x)]_{\rm av}$ at site $x$. (d) gives a contribution ${\cal A}(G_x) \propto {\rm Tr}\, O_x = 0$. (e) has ${\cal A}(G_x)\not=0$ but forces similar contractions on adjacent sites via the bond constraint, and these vanish as $q\to\infty$.}\label{2ptcon1}
\end{figure}

Contributions to the autocorrelation function ${\rm tr}\,[O(x,t) O(x)]$ are generated by contractions of the site diagrams shown in Fig.~\ref{2ptcon1}a and \ref{2ptcon1}b. The leading contribution $G_i$ at a site $i\not= x$ is from a contraction of the form shown in Fig.~\ref{2ptcon1}c. However, if this contraction is made at every site $i\not= x$, then because of the bond constraint it also applies at site $x$ and yields ${\cal A}(G_x) \propto {\rm Tr}\, O_x = 0$ in Fig.~\ref{2ptcon1}d. An example of an alternative contraction, for which ${\cal A}(G_x) \not= 0$, is shown in Fig.~\ref{2ptcon1}e. With such a choice at $i = x$, the bond constraint imposes contractions at nearby sites $i\not= x$ that are sub-leading in $q$.

We have also evaluated the leading non-zero contributions to autocorrelation function for large $q$ at small values of $t$. We find
\be
\ba
& 
\langle [O(x,t) O(x)]_{\rm av} \rangle= 	
\begin{cases}
	1 & $ for $t =0\, 
	\\
	0 & $ for $t =1\, .
	\\
	q^{-7}    & $ for $t =2 \, .
	\\
	16 q^{-11} &$  for $t =3 \, .
\end{cases}
\ea
\ee
These results are interesting for two reasons. First, the equivalent quantity for time evolution with a random unitary circuit  is identically zero at all $t\not= 0$. The results hence expose a difference between our Floquet model and random unitary circuits. The finite relaxation rate at finite $q$ in the Floquet model is consistent with expected generic behaviour, whereas complete relaxation for any $t\not= 0$ is likely to be a special feature of random unitary circuits. Second, and quite separately, the dominant contributions arise from non-Gaussian diagrams: for example, at $t=3$ the largest Gaussian term is $2q^{-9}$, but is cancelled by non-Gaussian contributions.

\subsection{Purity and R\'enyi entropies
}\label{ren}
\begin{figure*}[t]
	\centering
	\includegraphics[angle=0, width=1\linewidth]{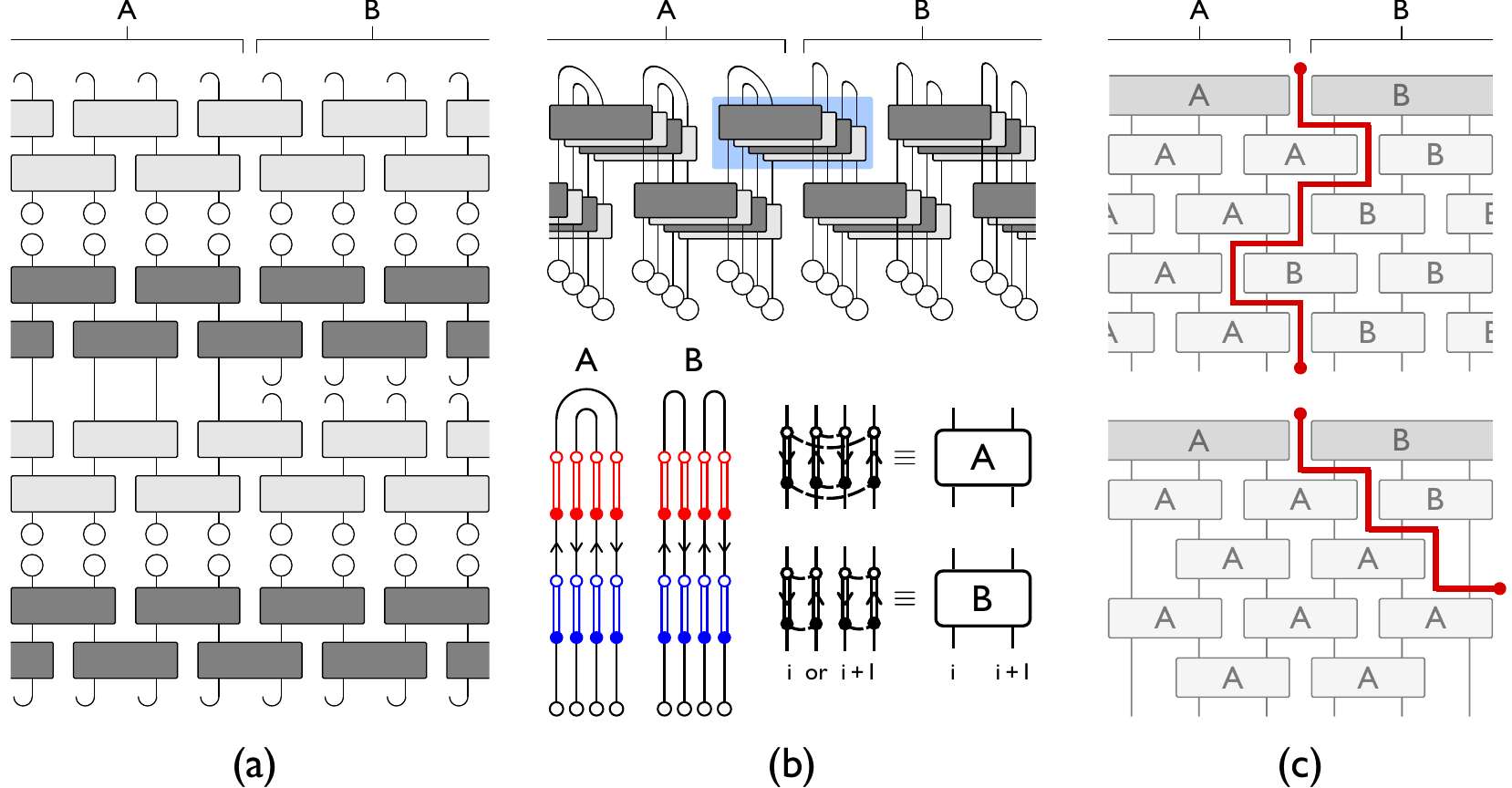}
	\caption{
\label{s2panel}
	Summary of the steps that convert the evaluation of the average purity into a combinatorial statistical mechanics problem.
	(a): Illustration of $e^{-S_2(t)}$ for $t=1$. Open circles represent a product state in the site basis and other conventions are as in Fig.~\ref{fig:model2}.
	(b) Top: Folded representation of $\exp[-S_2(t)]$ obtained by folding (a) so that timelines in $W(t)$ run upwards and those in $W^\dag(t)$ run downwards. 
	The large blue box highlights a stack of four unitary operators (see main text). 
(b) Bottom left: Site diagrams in regions $A$ and $B$. 
Note in particular 
how the timelines of sites in the regions $A$ and $B$ connect differently at the top of these diagram, because of the structure of traces in \eqref{renyidef}. 
(b) Bottom right: 
The two possible local contractions made within a block, which we label $A$ and $B$. Leading order diagrams only have such contractions. We represent these contractions in two alternative ways: either explicitly for a single site, as on the left side of the equivalences; or schematically for a pair of sites and a stack of unitaries, as on the right side of the equivalences. Note that the top boundary of the diagram can be treated as equivalent to blocks of such contractions. 
(c) Leading order diagrams for $\langle e^{-S_2(t)}\rangle$ at large $q$ are minimal-length domain wall (DW) diagrams. Their weight is $q^{-h_{AB}}$ with $h_{AB}$ being the length of horizontal DW  between $A$ and $B$-contractions.
The labels $A$ and $B$ on blocks in this figure indicate the type of local contraction.
 Top: A minimal-length DW diagram for $\langle \exp[-S_2(t)] \rangle$ at $t=4$ and $L>6$.  Bottom: One of the two minimal-length DW diagrams at $t>2$ and $L=6$. These two diagrams have the DW directed solely to the left or right. 
	}
\end{figure*}

We now show that the large-$q$ calculation of the purity can be reduced to counting domain wall (DW) configurations with positive weights, in a problem analogous to one from classical statistical mechanics. Related ideas apply (albeit more elaborately) to the evaluation of averages of the exponential of the  R\'enyi entropy $S_\alpha(t)$ for general positive integer $\alpha$ and we discuss the case $\alpha=3$ in Appendix \ref{3renapp}. Domain structure similar to what we derive here also appears in recent treatments of random unitary
circuits \cite{Nahum2017,Nahum2017a} and it is striking to find it as an emergent property of our Floquet model.


The main steps in our procedure are summarised in Fig.~\ref{s2panel}.
We wish to average $ \Tr_A \,[ \rho_A(t)^\alpha  ]$ for $\alpha=2$. 
Using the conventions of Fig.~\ref{fig:model2}, Eq.~\eqref{renyidef} has the pictorial representation shown in Fig.~\ref{s2panel}a.
Since each $\rho_A(t) = W^\dag(t) \ket{\psi} \bra{\psi} W(t)$, we have four sectors, containing $W(t), W^\dag(t), W(t)$ and $W^\dag(t)$ respectively. 
It is convenient to fold this diagram, as discussed in Sec.~\ref{diagrep} and 
{Fig.~\ref{fig:model2}. 
This procedure leads to a folded
representation containing four layers, as shown in Fig.~\ref{s2panel}b top. 
The folded site diagrams in regions $A$ and $B$ are shown in Fig.~\ref{s2panel}b bottom left. 
Note in particular 
how the timelines of sites in the regions $A$ and $B$ connect differently at the top of these diagrams, because of the structure of traces in \eqref{renyidef}. 
After averaging, it turns out that the leading contributions come from contractions of the $U$'s and $U^\dagger$'s that are brought to lie in a stack on top of each other by the folding operation. 
We call a contraction
of this type \emph{local}.
Such a stack of unitaries 
is indicated with a large blue box in Fig.~\ref{s2panel}b top.  In the evaluation of the purity (where two $U$'s and two $U^\dag$'s are involved) 
there are two possible local contractions (Fig.~\ref{s2panel}b bottom right). 

An intuitive explanation of the fact that only local contractions contribute for $q\to \infty$ is that any contraction between two dots in distant blocks would necessarily lead to a longer loop; as the total loop length is fixed, this implies a smaller number of loops and a lower order according to \eqref{mborder}.
This statement is proved in Appendix \ref{s2proof}. 


Given the restriction to local contractions, a further simplification of the diagrammatic representation is possible. It is no longer necessary to represent the unitaries within a stack individually. Instead we can simply depict the stack, using a label to indicate the type of local contraction. We label the two types of local contraction that appear in a calculation of the purity $A$ and $B$, because they involve the same types of pairing as are induced by the trace structure in regions $A$ and $B$ of the system. The diagrammatic notation applied to one stack is shown in Fig.~\ref{s2panel}b bottom right. The same notation is used for a larger system in Fig.~\ref{s2panel}c. 

We prove in Appendix \ref{s2proof} that the order of a diagram with only local contractions is given by $q^{- h_{AB}}$, where $h_{AB}$ is 
the number of segments of horizontal {wall} of length one lattice unit between an $A$ and a $B$ block. Such walls are shown with horizontal red lines in Fig.~\ref{s2panel}c. In this framework, the problem of finding the leading contributions admits a simple geometrical interpretation: they are represented by the set of minimal-length DW diagrams that separate the $A$ and $B$ contractions.
 Given the fixed boundary conditions for the blocks on the top of the diagram, a DW must connect the top centre to either the side or the bottom of the diagram  (Fig.~\ref{s2panel}c).

For $t \leq L/4$, a minimal-length DW can only connect the top centre to the bottom edge of the diagram. Since there are a total of $2t$ rows, the minimal $h_{AB}$ is $2t$, and the order is $q^{-2t}$. Moreover, there are two choices (for the DW to go left or right) below every row, so the number of leading order diagrams is $2^{2t}$.
For $t > L/4$, the lengths of the DW-s are minimized  when they connect the top centre and the side of the diagram with the shortest possible paths. This implies that there are only two leading order diagrams: those where the DW is directed exclusively towards either the left or the right. 
For these cases, $h_{AB} = L/2$ and the order is $q^{-L/2}$. 

In summary, Gaussian diagrams give the leading behaviour at large $q$:
\begin{equation} \label{reneq2}
\langle e^{-S_2(t)}\rangle  \sim \left\{
\begin{array}{cll}
2^{2t} q^{-2 t} & \quad & t\leq L/4 \\
2 q^{-L/2} & \quad & t > L /4
\end{array}\right.\;.
\end{equation}
Subleading terms eliminate the discontinuity in (\ref{reneq2}) at $t=L/4$. 
For larger $\alpha$, one can repeat the construction of Fig.~\ref{s2panel}. Each block has 
now $\alpha!$ possible states, corresponding to the possible local contraction inside a block of $\alpha$ $U$'s and $\alpha$ $U^\dag$'s. So, generically the problem reduces to counting, but for large $\alpha$ the procedure quickly becomes problematic. The case $\alpha = 3$ is discussed in Appendix \ref{3renapp}.

\subsection{Out-of-time-order correlator}

\begin{figure}[t!]
	\centering
	\includegraphics[angle=0, width=0.95\linewidth]{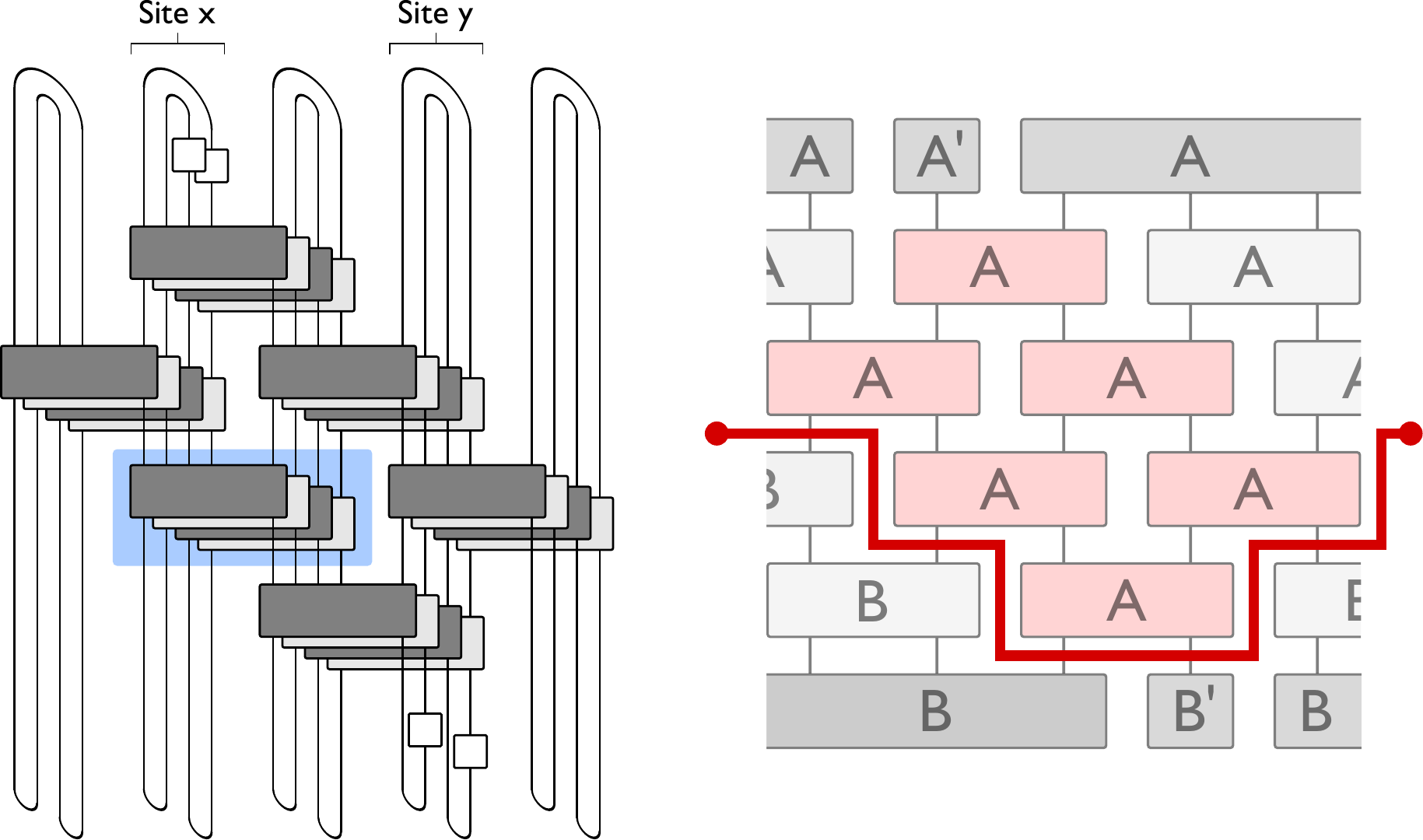}
	\caption{Diagrammatic representation of OTOC. Left: Folded diagram for ${\rm tr} [ O(x,t) O(y) O(x,t) O(y)]$ with $|x-y| =2$ and $t=2$. The blue box highlights four unitary operators,
	which can be represented in the block representation in a similar way to the ones in Fig.~\ref{s2panel}c. 
	Right: An apparent minimal-length horizontal DW diagram that vanishes due to $ {\rm tr}\, O(y) \equiv 0$. The red blocks are unitary pairs that can not be annihilated. The dark grey blocks represent the boundary conditions. $A'$ represents an effective contraction $A$ (Fig.~\ref{s2panel}c) with operators along the loops, and similarly for $B'$. 
	\label{otocfull}
	}\label{otoc1}
\end{figure}

Consider the diagrammatic representation of the OTOC defined in Eq.~\eqref{OTOCsimple}.
%
After the cancellation of $U$s with $U^\dagger$s wherever possible, we can distinguish two alternative scenarios. For  $t<  |x-y|/2$, all the unitary matrices are cancelled and we simply have
${\rm tr} [ O(x,t) O(y) O(x,t) O(y)] = {\rm tr} (O(x)^2) {\rm tr} (O(y)^2)  = 1$, independently of the realisation. 
Substituting the OTOC in \eqref{OTOCdef}, we see that the commutator vanishes. This  is consistent with the fact that the light-cone spreading from $x$ with velocity $v=2$ has not yet reached the operator in $y$. 
Conversely, for $t \geq |x-y|/2$ there is a region of unitary operators that cannot be cancelled. 

Calculation of the OTOC for $t \geq |x-y|/2$ is shown in Fig.~\ref{otoc1} left using a folded diagram.
The diagram contains four layers of $W$s, as for the purity, and we employ the block representation introduced in Sec.~\ref{ren}. 
Similarly to Sec.~\ref{ren}, it can be shown that the leading order diagrams necessarily involve only local contractions. 
Moreover, the trace structure of OTOC sets boundary conditions along the top of the diagram that act as $A$-blocks, and along the bottom that act as $B$-blocks. From this one would expect that the leading order diagrams are minimal-length DWs that cross the diagram horizontally, separating an upper domain of $A$s from a lower domain of $B$s (Fig.~\ref{otoc1} right). These diagrams however vanish, because they involve factors of either ${\rm tr}\, O(x) \equiv 0 $ or $ {\rm tr}\, O(y) \equiv 0$. In Appendix \ref{otocproof}, we prove that the leading contributions to the OTOC for $t \geq |x-y|/2$ go to zero for $q\to \infty$, and hence Eq.~\eqref{OTOC} is recovered in the large-$q$ limit.

\color{black}
\section{Summary and outlook}\label{sum}
In summary, we have introduced a minimal random-matrix model with extended spatial structure to study the
chaotic Floquet dynamics of a many-body quantum system.  We have presented a diagrammatic technique to compute several quantities using a systematic and controlled expansion in the inverse of the local Hilbert space dimension $q$.

Our study of a minimal model and the techniques we have developed are complementary to a variety of other recent works. In particular, the semiclassical approach to quantum chaos in few-body systems \cite{Haake} has been extended to bosonic systems at high density by considering interfering paths in Fock space that arise from solutions to the Gross-Pitaevskii equation, \cite{Richter2014,Richter2016} and to periodically driven spin systems in the large-spin limit. \cite{Guhr2017} These semiclassical techniques have the attraction of applying directly to specific physical systems that are of wide interest, rather than simply to a minimal model. They been used so far to identify particular many-body interference phenomena, but have not been developed to allow general computation of the dynamics of quantum information. In a separate advance, the Keldysh technique has been generalised to permit calculation of out-of-time order correlators, with applications to a variety of microscopic models. \cite{AleinerFaoroIoffe} These calculations are well-controlled in a quasiclassical regime, analogous to our large-$q$ limit, while the augmented Keldysh contour of Ref.~\onlinecite{AleinerFaoroIoffe} is (unsurprisingly) mirrored quite closely by the structure of the diagrammatic calculations we describe here. Further topical research,\cite{Gu2017} addressing a spatially extended version of the SYK model,\cite{kitaev2015simple} is set apart from the results we present by the fact that the zero-dimensional SYK model exhibits much greater structure than the individual random matrices of our model.

There are some obvious and interesting directions for additional investigations using the techniques we have set out. First, in contrast with random unitary circuits, our model has a well-defined Floquet operator, opening the possibility for the study of its spectral properties. It seems likely that further work in this direction will be useful, going beyond our evaluation of the spectral form factor in the random matrix regime. 
Second, the results we have presented are averages over an ensemble of systems. It would be useful to understand the magnitude of sample-to-sample fluctuations, by evaluating quantities such as $\langle (e^{-S_\alpha(t)})^2 \rangle - \langle e^{-S_\alpha(t)} \rangle^2$. More ambitiously, it would be appealing to use higher order terms in the $1/q$ expansion to search for the expected differences between the naive light-cone velocity, the entanglement-spreading velocity and the butterfly velocity, and to investigate broadening of the step in the OTOC given in (\ref{OTOC}) for the large-$q$ limit. 

There are also several generalisations. Our model and the techniques we have developed can be naturally 
extended to higher dimensions. Analogous models could also be developed for the other symmetry classes in random matrix theory.  In the context of quantum transport, it  would be interesting to incorporate the presence of conserved quantities, by modifying the local structure of the Floquet operator, as has been done recently for random quantum circuits. \cite{Huse2017,vonKeyserlingk2017a}

Finally, and speculatively, it is possible that an understanding of the ergodic phase in our model for chaotic many-body systems may play a role analogous to the treatment of a diffusive metal in the theory of disordered conductors, and provide a starting point for a theory~\cite{altland} of the many-body localisation transition.~\cite{basko2006metal}

After completing this work, we learned of recent related calculations of the spectral form factor for a many-body system, in which the equivalent of Eq.~(\ref{Ktdef}) is derived for the orthogonal  symmetry class~\cite{Prosen}.  There are however important differences between the two approaches. In particular, the result of [\onlinecite{Prosen}] develops 
from a specific long-range Hamiltonian using ideas of periodic orbit theory. Instead, the present work 
utilises a limit of large on-site Hilbert space and disorder averaging still retaining a local short-range structure. Therefore, one important potential use of our method is to
address ergodicity in disordered systems~\cite{paper2}.

\section*{Acknowledgements}

We thank Adam Nahum for extensive discussions. The work was supported in part by EPSRC Grant No. EP/N01930X/1.


\appendix
\section{Recursive formula for $V_{c_1, c_2, \dots , c_u}$} \label{recur}
The coefficient $V_{c_1, \dots , c_k}$ appearing in Eq.~(\ref{VVdef}) obeys the recursion relation \cite{Brouwer1996, Samuel, Mello}
\be
\ba
\delta_{c_1, 1}  & V_{c_2, \dots ,c_k}  = N  V_{c_1, \dots , c_k} 
+ \sum_{p+q = c_1} V_{p,q, c_2, \dots , c_k}
\\
& + \sum_{j=2}^{k}c_j V_{c_1 + c_j, c_2, \dots , c_{j-1}, c_{j + 1} , \dots, c_k}  \, . 
\ea
\ee
where $V_0 \equiv 1$ and $N=q^2$ is the dimension of the unitary group from which the Haar-distributed unitary operators are drawn.

\color{black}

\section{Enumeration of leading order diagrams}\label{orderdet}

In the following, we describe two methods for efficiently enumerating the leading order diagrams for the physical quantities we have computed. 
The contraction addition method described in App.~\ref{conadd} can be used to eliminate sub-leading single-site diagrams efficiently, which is particularly useful if the many-body diagram of interest has the same site-diagrams  across all sites (e.g. the spectral form factor).  
 The domain wall (DW) approach  described in App.~\ref{block} allows us to obtain an upper bound to the order, a global property of a diagram, by making only local calculations between neighbouring domains. 
 \color{black}
We apply the ideas from App.~\ref{conadd} in App.~\ref{levelcorr}, \ref{2ptproof} and \ref{otocproof}, and those from App.~\ref{block} in App.~\ref{s2proof} and \ref{otocproof}.

The foundation for both these methods is Eq.~\eqref{mborder}, which gives the order $\OO(\GG)$ of a diagram $\GG$ as a product of factors $q^{\tau_i + u_i - n_i}$ from each site. As the numbers $\tau_i$ of $T$-loops and $u_i$ of $U$-loops contribute to the order in the same way, we do not need to distinguish between $T$- and $U$-loops in the calculation of order discussed below. 


\subsection{Method of contraction-addition} \label{conadd}

An ensemble-averaged many-body diagram consists of $L$ site diagrams labelled by $i$, each with a number $n_i$ of contractions. The idea of this approach is that, 
given a site diagram, we can first remove all the contractions, and then
re-construct the diagram by adding contractions one at a time. The order can be evaluated by considering the effect of each contraction-addition.
The crucial point is that the procedure is independent of the sequence of contraction-additions. Each addition either leaves the order in $q$ unchanged (if it increases both $\tau_i+u_i$ and $n_i$ by one), or reduces the order (if the addition increases only $n_i$, or reduces $\tau_i+u_i$ and increases $n_i$). A convenient sequence (one that contains order-reducing contraction-additions in the first few steps) can therefore be used to eliminate sub-leading diagrams efficiently. 

To be more precise, given a diagram $\GG$, we consider the single-site configuration $\GG_i$ on site $i$. We choose a sequence of diagrams $\{\GG_{i}^{(0)},\ldots, \GG_{i}^{(n_i)}\}$, such that $\GG_i^{(0)}$ has no contractions,  $G_i^{(m+1)}$ is obtained from $G_i^{(m)}$ by adding one contraction, and $\GG_i^{(n_i)} = \GG_i$.
The overall order is then
\begin{equation}\label{contadd}
\OO(\GG) =   \prod_{i \in \text{site}} \OO(G_i^{(0)})   \prod_{m=1}^{n_i} \Delta_{i}^{(m)}\, . 
\end{equation}
Here, $\Delta_{i}^{(m)} = \OO(G_i^{(m)})/\OO(G_i^{(m-1)})$ is the change in the order when one contraction is added.

As we have established that we do not need to distinguish between $T$- and $U$-loops here, each contracted site diagram can be represented as a collection of loops, using the same notation for both types of loop. The same representation can also be used for each intermediate $G_i^{(m)}$ despite the fact that the notion of distinct $T$- and $U$-loops may not apply to $G_i^{(m)}$. The addition of a contraction has the effect either of merging two loops into a single longer one, or of breaking one loop into two. To see this, recall the prescription for construction of loops in a single-site configuration $\GG_i$ that is implied by Fig.~\ref{tloopuloop}. It can be expressed as follows. Start at any point on a single (for $T$-loops) or double (for $U$-loops) line; follow the direction along this line; when a dashed line is encountered, go to the other end of the dashed line and continue along the single or double line in the same sense; repeat until the starting point is reached: the path traced is a loop. Addition of an extra contraction to this prescription simply re-routes the paths of the two loop segments that meet the contraction from either side.
Moreover, since the two ends of the dashed line depicting a contraction are attached either to two separate loops or to two different portions of the same loop, we can
evaluate $\Delta_{i}^{(m)}$ omitting all information except what concerns the one or  two loops involved in the contraction. Once one focuses on the relevant details in this way, there are only eight distinct contraction-addition scenarios. We list them all in Fig.~\ref{appa1}, using a single unbroken line to denote generic loops. A minor complication is that it is necessary to distinguish loops that pass through one or two operators representing local observables from ones that do not, since these operators may also affect the weight of a diagram.


\begin{figure}[h]
	\centering
	\includegraphics[angle=0, width=1\linewidth]{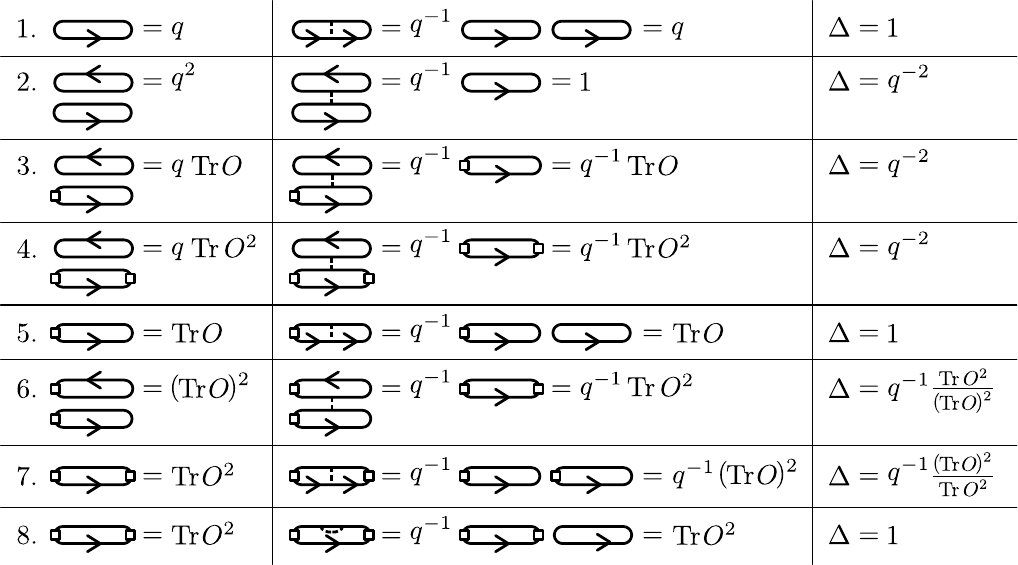}
	\caption{
	Enumeration of all distinct ways in which a contraction may be added to a site diagram. Columns, from left to right, are as follows. (i) Eight distinct combinations of one or two loops, before addition of a contraction, with associated contributions to ${\cal O}(G_i^{(m-1)})$. Three types of loop appear, containing zero, one or two local operators $O$ indicated using open squares. (ii) The same after adding a contraction, shown as a dashed line, and re-drawn purely as loops, with associated contributions to ${\cal O}(G_i^{(m)})$. (iii) Values of $\Delta^{(m)}$ arising from each contraction, obtained as the ratio of contributions in columns (ii) and (i). 
		}\label{appa1}
\end{figure}

We provide a simple example of how this approach to determining the order of a site diagram works in Fig.~\ref{appa2}.
\begin{figure}[h]
	\centering
	\includegraphics[angle=0, width=1\linewidth]{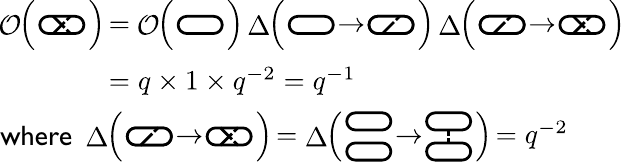}
	\caption{An example of order determination of a diagram using the method of contraction-addition. Arrows giving an orientation to loops are omitted for clarity.}\label{appa2}
\end{figure}


\subsection{Domain wall approach} \label{dwapproach}
\label{block}

In Sec.~\ref{ren} we set out a way of considering contributions to some quantities of interest, in terms of space-time domains within which contractions are all of the same type. The power of this approach lies in the fact that one can associate a bound on the diagram's order with each DW between two neighbouring domains.  This allows us to calculate a bound on the diagram's order, a global property, by making only local calculations between neighbouring domains. We explain details here, using
for concreteness the 2nd R\'enyi entropy (see Sec.~\ref{ren}) and referring to the block representation introduced in App.~\ref{AppB} and Fig.~\ref{blockrep}.

First, we rewrite the order of a diagram $\GG$
in terms of loops. 
We define the length $c_k$ of a loop $k$
as half of the number of non-dashed lines it contains. 
We make the convention that single lines at the top edge (where $W$ and $W^\dag$ are multiplied)
	are counted twice, so that lengths are always integer numbers (e.g.  at the top edge of Fig.~\ref{ex2} there are four single lines in each single-site diagram). This generalises the definition given in Sec. \ref{diagrep} to both $U$-loops and $T$-loops.
	We also define the total length $l_t = \sum_k c_k$ where the sum runs over all loops.
Then the order of a diagram in \eqref{mborder} can be re-written 
\begin{equation}
\label{orderloops}
\OO(\GG) =  q^{l_{\text{t}} -n}  
\prod_{k \in \text{loops}}  \Gamma_{c_k} \, ,
\end{equation}
where $\Gamma_c = q^{1-c}$ and $n = \sum_i n_i$
is the total number of contractions.
The values of $l_t$ and $n$ are fixed by the observable of interest. In particular, for the 2nd R\'enyi entropy $l_t = n$.

The intuition behind Eq.~\eqref{orderloops} is as follows. Since a given diagram has a fixed number of length segments, its order is high (low) when there are many short (a few long) loops. The pre-factor $q^{l_t -n}$ represents the theoretical maximum order when all length segments are used to form 1-loops. $\Gamma_c$ is the cost associated with a loop of length $c$, and its value decreases as $c$ increases. For example, $\Gamma_1$ has value 1 since 1-loop is the shortest possible loop. $\Gamma_2$ has value $q^{-1}$ because one could have formed two $1$-loops (giving a factor of $q^2$ in Eq.~\eqref{mborder}) instead of one $2$-loop (giving a factor of $q$). 


Next, we rewrite the order of a diagram in terms of boundaries that intersect the loops. We define a \textit{boundary} as a \textit{horizontal} segment of unit lattice spacing that separates two blocks (for example, the red lines in Fig.~\ref{appb1} top), so that each block has four boundaries (except for the blocks representing the trace structure at the upper edge, e.g. zero-th row in Fig.~\ref{appb1} top). Every loop instersects two such boundaries per unit length. To each boundary $w$, we can therefore associate a cost 
\begin{equation}\label{wallcost}
\mathcal{C}(w) = \prod_{\substack{k \in \text{loops} \\ k \nmid w}} \Gamma_{c_k}^{1/2c_k}
\end{equation}
where $k$ labels the loops that the boundary $w$ intersects, and $c_k$ is the corresponding loop length. In this way, the order of a diagram $\GG$ is expressed in terms of the cost of each boundary, so that
\be \label{wallorder1}
\OO(\GG)
=   q^{l_{\text{t}} -n}  \prod_{w \in \text{walls}} \mathcal{C}(w) \;.
\ee
Since $\Gamma_1 = 1$, loops of unit length do not contribute to $\OO(\GG)$. We obtain an upper limit
on the cost of a boundary by assuming longer loops to have length not more than $c_k = 2$. This leads to
\begin{equation}
\label{wallorderbound}
\mathcal{C}(w) = \prod_{\substack{k \nmid w, c_k \geq 2}} \Gamma_{c_k}^{1/2c_k} \leq q^{-n_w/4}
\end{equation}
where $n_w$ is the number of loops crossing the boundary $w$ with length longer than $1$. 
Referring to examples in Fig.~\ref{appb1}, $\mathcal{C}(w)$ for the top sub-figure is bounded by (and equal to) $q^{-1}$ with $n_w =4$, and $\mathcal{C}(w)$ for the bottom sub-figure is bounded by $q^{-3/4}$ with $n_w =3$. As we will see, a costly boundary (one with $n_w > 0$) is always sandwiched between blocks with different contractions. It is therefore natural to call such a boundary a \textit{domain wall} (DW). With minor modifications, this method can also be applied to diagrams that contain local operators.


\color{black}
\section{Spectral form factor}\label{levelcorr}

Here 
we use the method of contraction-addition (App.~\ref{conadd}) to enumerate the leading order diagrams for $\langle \levc{t} \rangle$. In Eq. \eqref{contadd} we have $n_i = 4t$ and $\mathcal{O}(G_i^{(0)}) = q^2$. There are only two relevant contraction-addition scenarios illustrated in Fig.~\ref{appa1}: (i) An intra-loop addition (item 1 in figure), where the two legs of the new contraction line land on the same loop. 
Then $\Delta= q^{0}$ according to Eq.~\eqref{mborder} because $\tau +u$ and $n$ both increase by one. (ii) An inter-loop addition (item 2 in figure), where the two contraction legs land on two different bare loops. Since the two loops merge into a single bigger loop due to the new contraction, $\tau +u$ decreases by one, and $n$ increases by one. So  $\Delta=q^{-2}$. 


The many-body diagram for $\levc{t}$ comprises site diagrams before contraction as in Fig.~\ref{level1}a. For $t=0$ we have $\levc{t}=   q^{2L}$ since $W(0) = \openone_{q^L}$. 
To compute the large--$q$ limit of $\levc{t}$ for $t>0$, we first note that there are diagrams with multiple contractions that are $O(1)$, such as Fig.~\ref{2ptcon1}c.  
This is in fact a highest order diagram for $t>0$ according to Eq.~\eqref{contadd}, because there must be at least one inter-loop contraction-addition which is associated with  $\Delta=q^{-2}$ per site, and because, from Fig.~\ref{appa1}, the later contraction-additions can be made without increasing the order of the diagram. So any diagrams with order smaller than $O(1)$ are sub-leading. 

There are $t$ leading order site diagrams on site $i$, by the following argument. Using the fact that the order determination is independent of the sequence of contraction-addition, we choose to contract on site $i$ the (filled and un-filled) blue dots on the top layer from left to right. The first filled blue dot can be contracted with any one of the $t$ filled blue dots on the bottom layer. This inter-loop contraction costs $q^{-2}$. In order to obtain a leading diagram, the later contractions must all be of  $O(1)$, i.e. intra-loop contractions. A general feature of an $O(1)$ contraction is that
it must partition a bigger loop into two smaller loops, such that the $q$ factor associated with the extra loop (since $\tau$ or $u$ increases by one) cancels the $q^{-1}$ factor of the contraction (since $n$ increases by one) (see Fig.~\ref{appa1}). Furthermore, on each of the two smaller loops, there must be equal numbers of un-contracted blue dots on the top and bottom layer. Otherwise, there will be an inter-loop contraction which will render the diagram sub-leading. It is straightforward to see that there is a unique choice of contraction of the second blue dot that satisfies this requirement. Similarly, there are unique choices for the rest of the blue dots on this site. In order to not incur further cost on site $i$, each red dot on the top layer must be contracted with the only other red dot on the same loop on the bottom layer. So there are unique choices for the red dots as well.

Due to the bond constraint, the $(i+1)$-th site diagrams inherit the choice of either the blue or red dot contractions on site $i$. We can repeat the above analysis site-by-site and conclude that there are $t$ leading order diagrams of $O(1)$ for $t>0$ (Fig.~\ref{level1}b,c and d). 

Lastly, each of these $t$ diagrams is translated algebraically to the factor $1$, simply observing that it is Gaussian and for each site $\tau_i + u_i = n_i$ in \eqref{mborder}.
In other words, for $t>0$, $\levc{t}= t $, and we have arrived at Eq. \ref{Ktdef}. 

\section{Relaxation of local observables}\label{2ptproof}
Here we prove Eq.~\eqref{2pt}. For $t=0$, the calculation is straightforward since $W(0) = \openone_{q^L}$. For $t=1$, there is only one configuration at each site because there is only one pair of dots of each kind. Since we have chosen $\Tr O_x =0$, the algebraic term associated with this diagram vanishes. For $t>1$, we use the same argument as in App.~\ref{levelcorr} to show that every possible diagram has an order that vanishes as $q \to \infty$.  In the formulation of contraction-addition (App.~\ref{conadd}), we have $n_i = 4t$ and $\mathcal{O}(G_i^{(0)})=1$. On site $x$, we choose to first contract the blue dot on the bottom left of the diagram (Fig.~\ref{2ptcon1}b). This bottom left blue dot cannot be contracted with the bottom right blue dot non-trivially since $\Tr O_x$ is zero. However, if this bottom left blue dot is contracted with any other blue dots, it will partition a bigger loop into two smaller loops, each of which has an unequal number of dots from $U$s and $U^\dagger$s. This implies that there will be at least one addition of an inter-loop contraction with $\Delta \propto q^{-1}$ (item 6 in Fig.~\ref{appa1}). Since all other contraction-additions can only further reduce the order according to Fig.~\ref{appa1}, we have $\mathcal{O}(G) < q^{-1}$ for any diagram $G$ associated with $\langle [O(x,t)O(x)]_{\rm av} \rangle$ when $t>1$. 

\section{2nd R\'enyi entropy}
\label{s2proof}

The diagramatic representation used in Fig.~\ref{s2panel}c requires a straightforward extension when it is empolyed in the presentation of detailed proofs, and we start by describing this extension. It arises for the following reason. The rectangles of Fig.~\ref{s2panel}c represent local contractions for a stack of two $U$'s and two $U^\dagger$'s. Each contraction involves linking both open dots with open dots, and closed dots with closed dots. To allow for the possibility that different pairings are made for open dots and for closed dots, each rectangle appearing in Fig.~\ref{s2panel}c is divided horizontally into two in this Appendix.

\subsection{Block representation of $\langle \exp[-S_2(t)]\rangle$}\label{AppB}

\begin{figure}[h]
	\centering
	\includegraphics[angle=0, width=1\linewidth]{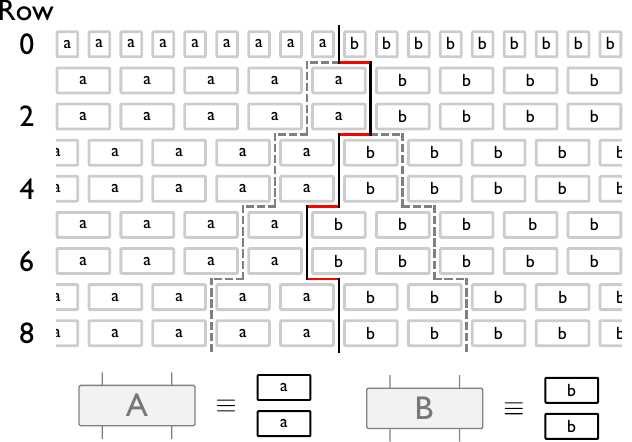}
	\caption{Top: The block representation of a leading DW diagram for  $\langle \exp[-S_2(t)]\rangle$. The top row (the zero-th row) denotes the effective blocks that represent the structure of traces in sub-regions $A$ and $B$ (Fig.~\ref{s2panel}b bottom left). The dashed line represents a light cone outside of which all unitary pairs can be annihilated as described in Fig.~\ref{fig:model2}. The red line represents costly boundaries (between blocks of different contraction types, see Appendix~\ref{dwapproach} and  Table~\ref{dwcost}). Bottom: The correspondence between blocks and the symbols used in the bottom right of Fig.~\ref{s2panel}b. 
	}\label{blockrep}
\end{figure}
\begin{figure}[h]
	\centering
	\includegraphics[angle=0, width=1\linewidth]{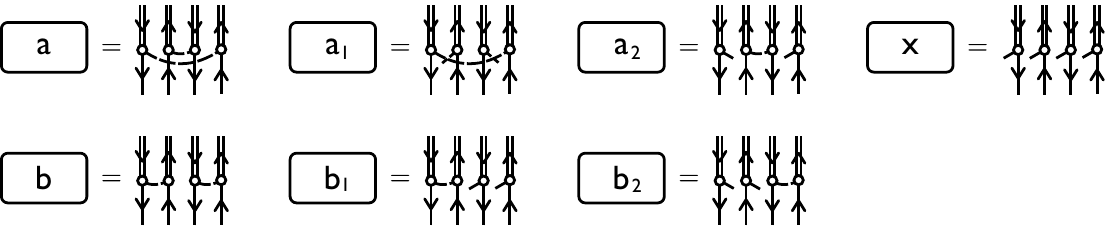}
	\caption{Contractions in a portion of a site diagram for $\langle \exp[-S_2(t)]\rangle$, showing all seven possible block types.
	}\label{block1alt}
\end{figure}

In detail, we introduce the \textit{block representation} as follows. Beginning from the folded representation introduced in Sec.~\ref{ren}, we consider a stack of four unitary operators (blue box in Fig.~\ref{s2panel}). We use Fig.~\ref{diag1} to represent each unitary operator in terms of dots and double lines. A \textit{block} is defined as a region within this box that encloses only filled (or only empty) dots. In this way, to each stack of four unitaries, we associate two blocks containing respectively filled and empty dots (so that, the four-legged symbol in the bottom right of Fig.~\ref{s2panel}b corresponds to two blocks in Fig.~\ref{blockrep} bottom). For $\langle \exp[-S_\alpha(t)]\rangle$, there are $4t$ rows of such blocks, and drawing each of them as an empty rectangle leads to the representation in Fig.~\ref{blockrep}.

\color{black}
We categorize the possible contractions of the dots within a block into seven types $\mathcal{T} = \{a,b,a_1,a_2,b_1,b_2,x\}$. As explained in the main text, $a$ and $b$ only involve local contractions 
(within the block), while $x$ involves only non-local contractions. The other types are of mixed local and non-local character, and are defined in Fig.~\ref{block1alt}. 
An upper bound $\omega(c,c')$ for the cost of the boundary between two neighbouring block configurations $c, c' \in \mathcal{T}$ is shown in Table \ref{table1}. Two examples of the evaluation of $\omega (c, c')$ are illustrated in Fig.~\ref{appb1} using the method introduced in Appendix~\ref{dwapproach}.
\begin{figure}[h]
	\centering
	\includegraphics[angle=0, width=1 \linewidth]{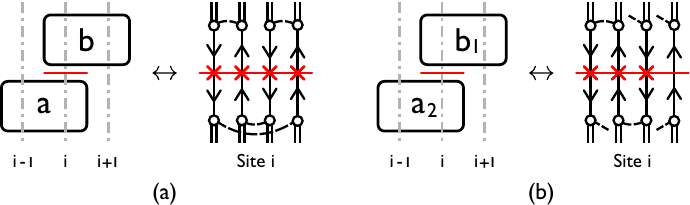}
	\caption{Examples of contractions in a portion of a site diagram for $\langle \exp[-S_2(t)]\rangle$. Left sides of (a) and (b): 
	 two neighbouring blocks along site $i$, shown in the space-time plane with the time axis vertical. 
	 Right sides of (a)  and (b): equivalent site diagrams with the time axis vertical; open-ended dashed lines are contracted non-locally with an external block. Red lines represent sections of boundary (of unit length in the left-hand pictures).  (a) right: the boundary crosses four times a loop of length at least $2$ at the red crosses, so $\omega(a, b) = q^{-1}$. (b) right: The boundary has three costly crossings, so $\omega(b_1, a_2) = q^{-3/4}$.}\label{appb1}
\end{figure}
\begin{table}[h]
	\begin{tabular}{ l | l | l | l | l | l | l | l} \label{table1}
		$\omega$ &  $a$ & $b$ & $a_1$ & $a_2$ & $b_1$ & $b_2$ & $x$\\
		\hline
		%
		$a$     &  $q^{0}$ & $q^{-1}$ & $q^{-1/2}$ & $q^{-1/2}$ & $q^{-1}$ & $q^{-1}$ & $q^{-1}$\\
		$b$     &               & $q^{0}$ & $q^{-1}$ & $q^{-1}$ & $q^{-1/2}$ & $q^{-1/2}$ & $q^{-1}$\\
		$a_1$ &               &              & $q^{0}$ & $q^{-1}$ & $q^{-3/4}$ & $q^{-3/4}$ & $q^{-1/2}$\\
		$a_2$ &               &              &              & $q^{0}$ & $q^{-3/4}$ & $q^{-3/4}$ & $q^{-1/2}$\\
		$b_1$ &               &              &              &              & $q^{0}$ & $q^{-1}$ & $q^{-1/2}$\\
		$b_2$ &               &              &              &              &              & $q^{0}$ & $q^{-1/2}$\\
		$x$     &               &              &              &              &              &              & $q^{0}$\\
	\end{tabular}
	\caption{Upper bounds for the cost associated with the boundaries between the seven possible types of block contraction.
		The matrix is symmetric and so only the upper triangle is written explicitly. 
		\label{dwcost}}
\end{table}

\subsection{Evaluation of $\langle \exp[-S_2(t)]\rangle$}\label{s2app}


We discuss diagrams for $\langle \exp[-S_2(t)]\rangle$ in terms of different block pairings.
We first fix our convention as follows. We label the rows of blocks from the top as in Fig.~\ref{blockrep}.  
We refer to the boundaries immediately above the $p$-th row of blocks as
the $p$-th row of boundaries. 

Due to the simplification illustrated in Fig.~\ref{fig:model2}, pairs of $U$ and $U^\dagger$ can be cancelled outside of the light cone originating from the sub-system boundary (shown with dashed lines in Fig.~\ref{blockrep}). Alternatively, instead of cancelling them, it is equivalent to assume that the blocks on the far left and far right have respectively $a$- and $b$- contractions, as long as we remain in the large-$q$ limit. Consistently, this choice implies costless boundaries according to Table~\ref{dwcost}.
Therefore, in particular, a configuration for any given odd row of boundaries can be parametrised as in Fig.~\ref{wallrow}, and we denote this as $(a d_1 d_2 d_3 \dots d_k b)_{\text{wall}}$ with variables $d_1 \neq a$, $d_k \neq b$, and $d_i \neq d_{i+1}$ for $i=1, \dots , k-1$. $a$, $b$ and $d_i$ represent (connected) domains of blocks with the same type of contractions. Note that for every change of domain, there is a DW associated with a factor of $q^{-m}$ with $m>0$. 

\begin{figure}[h]
	\centering
	\includegraphics[angle=0, width=1 \linewidth]{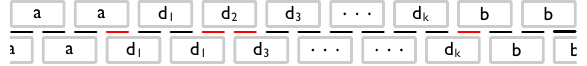}
	\caption{A generic configuration of an odd row of boundaries denoted as $(a d_1 d_2 d_3 \dots d_k b)_{\text{wall}}$. Costly boundaries (DWs) are drawn in red. Costless boundaries are drawn in black.}\label{wallrow}
\end{figure}

We claim the following for the leading diagrams. 
(i) Each odd row of walls has the form $(a b)_{\text{wall}}$.
(ii) Each even row of walls is sandwiched between two rows of blocks with identical configurations.
(iii) The leading diagrams have order $q^{-2t}$.

An example of order $q^{-2t}$ can be found by constructing a single DW dividing two domains of blocks with contraction $a$ and $b$ (Fig.~\ref{blockrep}), such that on the odd rows of boundaries, the order is $\omega(a,b) = q^{-1}$, and on the even rows of boundaries 
 the order is $1$. So any diagram with order lower than $q^{-2t}$ is sub-leading.

To prove these statements, we note that an odd row $(a b)_{\text{wall}}$ has only one DW segment with cost $q^{-1}$. 
Conversely, by using the costs in Table~\ref{dwcost}, it is easy to verify that
any row configuration $(a d_1 d_2 d_3 \dots d_k b)_{\text{wall}}$ with $k\geq 1$ always has an order strictly smaller than $q^{-1}$. Note that the order of a row with $k\geq 2$ is straightforwardly smaller than $q^{-1}$ since each wall costs at least $q^{-1/2}$.


An even row of walls that is sandwiched by two rows of blocks with different configurations is associated with an order of $q$ lower than $1$ because there is a costly wall sandwiched by two blocks of different types  (see Appendix~\ref{AppB}). 

Leading diagrams are obtained maximising the order on odd and even rows, thus proving (i) and (ii). 
%
Statement (iii) follows trivially from statements (i) and (ii).  Finally, (ii) implies that these diagrams are Gaussian (see Sec.~\ref{diagrep}) and translate into positive algebraic factor. Hence, we have recovered Eq.~\ref{reneq2}.

\color{black}

\section{3rd and higher R\'{e}nyi Entropies}\label{3renapp}
\begin{figure}[h]
	\centering
	\includegraphics[angle=0, width=0.8\linewidth]{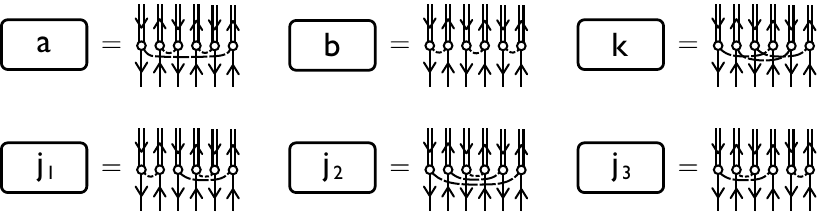}
	\caption{The six local contractions for a block appearing in  $	\langle \exp[-S_3(t)]\rangle$ with the same notation as the one in Fig.~\ref{block1alt}.}\label{3ren1}
\end{figure} 
We can also represent the leading contributions to $\langle \exp[-2 S_3(t)]\rangle$ by using a block representation . An extra feature in this case 
compared with $\alpha=2$ is that there are six possible types of local contraction within a block. We label these $a$, $b$, $k$, and $j_i$ for $i=1,2,3$ (Fig.~\ref{3ren1}). The blocks at the top edge of the diagram are fixed to be of type $a$ and $b$ by the structure of traces in $\exp[-S_3(t)]$.

Repeating the approach for $\alpha=2$, we find that leading diagrams only involve local contractions.
Equivalent statements to (i), (ii) and (iii) in Appendix~\ref{s2app} apply. For (i), we find that the leading order odd rows of boundaries have order $q^{-2}$, and they are of the forms $(ab)_{\text{wall}}$ or $(a j_i b)_{\text{wall}}$. All other odd row configurations  $(a d_1 d_2 d_3 \dots d_k b)_{\text{wall}}$ are associated with factors smaller than $q^{-2}$. For $k> 4$, this is trivial because each wall costs at least $q^{-1/2}$. For $ 1 \leq k \leq 4$, we enumerate all cases using symbolic computation to complete the proof. Statement (ii) for $\alpha =3$ is the same as the one for $\alpha =2$, and (iii) follows trivially for time $t \leq L/4$.
 An example of a leading order [$O(q^{-4t})$] diagram is given in Fig.~\ref{3ren3}.


The main difference with respect to $\alpha=2$ is that the leading contributions also include diagrams with regions of $j_i$-contractions, as well as regions of $a$- and $b$-contractions. We classify the leading diagrams at $t$ by the width $r$ of the $j_i$-region on the bottom row of blocks. 
To count the number of leading diagrams, we use an inductive approach from time $t$ to $t+1$. 
Every leading diagram at $t+1$ can be generated from a leading diagram at $t$ by adding four rows of blocks at the bottom of the diagram at $t$. A newly-added even row must have a configuration identical to the one above. A newly-added odd row has configuration depending on the one above according to the following rules (Fig.~\ref{3ren2}).  
For each even row in which the width of the $j_i$-region is zero, there are five possible configurations for the odd row below: two with width 0 and three with width 1.  For each even row with width $r>0$, there are four possible configurations for the odd row below, two with width $r$, one with width $(r-1)$ and one width $(r+1)$. An example of a leading diagram is given in Fig.~\ref{3ren3}.
\begin{figure}[h]
	\centering
	\includegraphics[angle=0, width=1\linewidth]{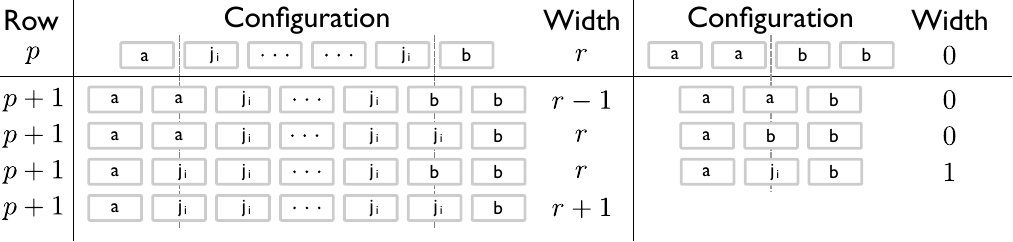}
	\caption{
		The recursive rules for enumerating
		the $p+1$-th row of the leading order diagrams given the $p$-th row of the leading order diagram.			
		These rules are derived from the fact that, in a leading diagram, the difference between the DW positions at even row $p$ and odd row $p+1$ is exactly one lattice spacing.}\label{3ren2}
\end{figure} 

We can write this recurrence relation in matrix form, by introducing a vector $\mathbf{v}(p) = (v_0 (p) , v_1 (p) , \dots)^T$ where $v_r (p) $ denotes the number of diagrams with total number of rows $p$  that have a $j_i$-region of width $r$ in the last row. 
The degeneracy $D(t)$ can be obtained by summing up the components of $\mathbf{v}(t)$. These in turn can be found by acting $2t$ times with a transfer matrix representing the recursive rules on the vector $\mathbf{v}(t=0) = (1,0,0, \dots)^T$. 
\be
D(t) = 
\begin{bmatrix}
	1 & 1 & \Cdots  
\end{bmatrix}	
\begin{bmatrix}
	2    & 1   &         &     &      \\
	3    & 2   & 1       &   &            \\
	& 1   & 2   &  1    &     \\
	&    &  1  &   2 &       \Ddots  \\
	&           &        &  \Ddots    & \Ddots     \\
\end{bmatrix}^{2t}
\begin{bmatrix}
	1  \\
	0 \\
	0         \\
	\Vdots              \\
	\phantom{\Vdots}
\end{bmatrix}
\sim \left(\frac{4 + 3 \sqrt{2}}{2} \right)^{2t} \, .
\ee 
\color{black}
On the right side of this expression we have used the fact that, for $1 \ll t < L/4$, $D(t)$ is dominated by the largest eigenvalue of the transfer matrix.

For $t > L/4$, there are five leading order diagrams. The first two have $ab$ DW directed only towards the left or the right. The other three diagrams have only $j_i$-contractions within the light-cone region (dashed lines in Fig.~\ref{3ren3}), with the $a j_i$ and $j_i b$ DW-s directed only towards the left and the right respectively
After translating these diagrams into algebraic terms, we obtain \eqref{reneq} for $\alpha =3$.

For higher R\'{e}nyi entropies,  we can prove for $\alpha =4$ that leading diagrams consist of local contractions only, and that $K_4 = \text{Cat}(4)$ in Eq.~\ref{reneq}. If we assume the analogous statements about local contractions for $\alpha > 4$, we can apply the above approach and show that $K_\alpha = \text{Cat}(\alpha)$ up to $\alpha = 10$.

\begin{figure}[h]
	\centering
	\includegraphics[angle=0, width=1\linewidth]{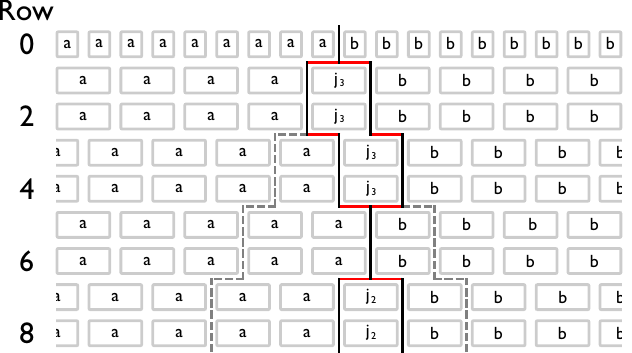}
	\caption{
		A leading order DW diagram in the block representation for $\langle \exp[-2 S_3(t)]\rangle$ in the large--$q$ limit.
		The order is given by $q^{-2 h_{ab} - h_{aj_i} - h_{j_i b}}$ where $h_{c_1 c_2}$ counts the horizontal walls between $c_1$- and $c_2$-contractions in red.	
	}\label{3ren3}
\end{figure}

\section{Evaluation for OTOC}
\label{otocproof}

Here we prove that the OTOC, $\langle {\rm tr} [O(x,t) O(y) O(x,t) O(y)] \rangle$, vanishes at large $q$ for $t \geq |x-y|/2$.
We employ the analogous block representation introduced in App.~\ref{AppB} of the main text. 
Using the contraction-addition method described in 
Appendix~\ref{orderdet}, the order of a diagram can be written as in Eq. \eqref{contadd} with $\mathcal{O}(G_i^{(0)}) = 1$.

We choose to contract blocks of $U$s and $U^{\dagger}$s bond-by-bond from the left to the right in Fig.~\ref{otoc1} . For the left-most bond, there are always two blocks of filled and empty dots for all $t \geq x/2$. For each of these blocks there are two choices of contraction: $A$ or $B$.
A choice of contraction-$A$ in one of the two blocks can be followed by other contractions in two alternative ways. Either all other blocks within its upward light cone (in this case a stripe of blocks) have $A$-contractions: this results in a factor $\propto (\Tr O)^2 =0$ at site $x$, due to contraction-addition of type seven in Fig.~\ref{appa1}.  
Alternatively there is a $q^{-1}$ cost due to at least one inter-loop contraction-addition of type two, three or six in Fig.~\ref{appa1}.
Similarly, the choice of contraction-$B$ would give a vanishing contribution because of the operator in $y$.
These arguments are illustrated in Fig.~\ref{otoc2}.
So we have shown that all diagrams have an order bounded by $q^{-1}$ for $t \geq |x-y|/2$. By contrast, for $t < |x-y|/2$ all $U$s and $U^\dagger$s can be cancelled, leaving only $\langle {\rm tr} [O(x)^2]  {\rm tr} [O(x)^2] \rangle =1$.


\begin{figure}[h]
	\centering
	\includegraphics[angle=0, width=0.75\linewidth]{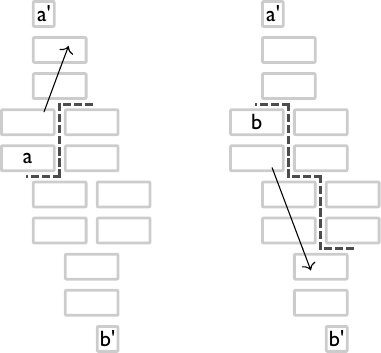}
	\caption{A partially contracted diagram for the OTOC at $t=2$ in the block representation. The choice of an $A$-contraction (or a $B$-contraction) of blocks on the left-most bond forces blocks in the upper (lower) light cone (dashed lines) to have the same type of contraction. Otherwise, the diagram has an order bounded by $q^{-1}$.}\label{otoc2}
\end{figure}

\bibliography{FloquetChaos}
\end{document}